\documentclass[pra,twocolumn,showpacs,superscriptaddress,floatfix,aps,10pt]{revtex4-1}
\usepackage{graphicx}
\usepackage{amssymb,amsmath,latexsym,bm,dsfont}
\usepackage[shortcuts]{extdash} 
\usepackage{verbatim}

\renewcommand{\vec}[1]{\bm{\mathrm{#1}}}
\newcommand{\vhat}[1]{\hat{\bm{\mathrm{#1}}}}
\makeatletter\def\@pacs@name{DOI: }\makeatother 
\begin{document}
\title{Asymmetric skyrmion Hall effect in systems with a hybrid Dzyaloshinskii-Moriya interaction}
\author{Kyoung-Whan~Kim}
\email{kyokim@uni-mainz.de}
\affiliation{Institute of Physics, Johannes Gutenberg University Mainz, 55099 Mainz, Germany}%
\author{Kyoung-Woong~Moon}%
\affiliation{Spin Convergence Research Team, Korea Research Institute of Standards and Science, Daejeon 34113, Republic of Korea}
\author{Nico~Kerber}%
\affiliation{Institute of Physics, Johannes Gutenberg University Mainz, 55099 Mainz, Germany}%
\affiliation{Graduate School of Excellence Materials Science in Mainz, 55128 Mainz, Germany}%
\author{Jonas~Nothhelfer}%
\affiliation{Institute of Physics, Johannes Gutenberg University Mainz, 55099 Mainz, Germany}%
\author{Karin~Everschor-Sitte}%
\email{kaeversc@uni-mainz.de}%
\affiliation{Institute of Physics, Johannes Gutenberg University Mainz, 55099 Mainz, Germany}%
\date{\today}

\begin{abstract}
We examine the current-induced dynamics of a skyrmion that is subject to both structural and bulk inversion asymmetry. There arises a hybrid type of Dzyaloshinskii-Moriya interaction (DMI) which is in the form of a mixture of interfacial and bulk DMIs. Examples include crystals with symmetry classes C$_n$ as well as magnetic multilayers composed of a ferromagnet with a noncentrosymmetric crystal and a nonmagnet with strong spin-orbit coupling. As a striking result, we find that, in systems with a hybrid DMI, the spin-orbit-torque-induced skyrmion Hall angle is asymmetric for the two different skyrmion polarities ($\pm 1$ given by out-of-plane core magnetization), even allowing one of them to be tuned to zero. We propose several experimental ways to achieve the necessary straight skyrmion motion (with zero Hall angle) for racetrack memories, even without antiferromagnetic interactions or any interaction with another magnet. Our results can be understood within a simple picture by using a global spin rotation which maps the hybrid DMI model to an effective model containing purely interfacial DMI. The formalism directly reveals the effective spin torque and effective current that result in qualitatively different dynamics. Our work provides a way to utilize symmetry breaking to eliminate detrimental phenomena as hybrid DMI eliminates the skyrmion Hall angle.
\end{abstract}

\maketitle

\section{Introduction}

Magnetic skyrmions~\cite{Bogdanov1989,Bogdanov1989b,Bogdanov1999} are localized magnetic textures with nontrivial topology in real space and have been experimentally realized as lattice structures~\cite{Muhlbauer2009a,Yu2010} and individual entities~\cite{Romming2013}. They have received much attention due to interesting phenomenology arising from their topological nature, such as the skyrmion Hall effect due to the Magnus force~\cite{Everschor2011, Iwasaki2013a,Jiang2016,Litzius2017} and the topological Hall effect due to the emergent electromagnetic fields~\cite{Bruno2004,Neubauer2009,Schulz2012,Everschor2014}.
Stabilization of highly localized skyrmions is achieved by the Dzyaloshinskii-Moriya interaction (DMI)~\cite{Dzyaloshinskii1957,Moriya1960b,Fert1980}. The DMI is an antisymmetric exchange interaction that breaks the chiral symmetry of magnetic textures and favors skyrmions with a particular chirality. There are two types of DMI mainly considered in the literature, commonly denoted as the interfacial and the bulk DMIs. The interfacial DMI arises from structural inversion asymmetry, typically present at interfaces between a ferromagnet and an adjacent layer with strong spin-orbit coupling~\cite{Fert1980,Kim2013b,Yang2015b}.
The bulk DMI arises from noncentrosymmetric crystal structures typically observed in B20 compounds~\cite{Pfleiderer2010} such as MnSi~\cite{Muhlbauer2009a}, Fe$_{1-x}$Co$_x$Si~\cite{Yu2010}, FeGe~\cite{Yu2011b}, and Mn$_{1-x}$Fe$_x$Ge~\cite{Shibata2013,Grigoriev2013,Gayles2015,Turgut2018,Koretsune2015}.
In systems with strong DMI, the interfacial DMI usually stabilizes N\'{e}el skyrmions~\cite{Jiang2015a,Woo2016,Boulle2016} while the bulk DMI in B20 compounds stabilizes Bloch skyrmions~\cite{Muhlbauer2009a,Yu2010}.
Although more general crystal symmetries allow for generalized forms of bulk DMI~\cite{Birss1964, Bogdanov1989,Kezsmarki2015,Hoffmann2017}, within this paper, we denote the one stabilizing Bloch skyrmions by the bulk DMI~\cite{Landau1984}.

Skyrmions are considered as promising candidates for spintronic applications like racetrack memories~\cite{Parkin2008, Fert2013}.
Advantages are the low critical current density to depin skyrmions in a nanowire~\cite{Jonietz2010,Schulz2012}, their tendency of being less sensitive to impurities~\cite{Iwasaki2013}, and the recent discovery of low pinning materials with the skyrmion motion~\cite{Zazvorka2018}. While being an interesting consequence originating from topology, the skyrmion Hall effect imposes an outstanding challenge for applications.
It drives a transverse motion of a skyrmion to the applied current, pushing it to the boundary of a nanowire 
and possibly annihilating it for typical current density used for racetrack memories~\cite{Iwasaki2013a}. Hence, optimal for fast driving speeds and information delivery rate is a straight skyrmion motion along the nanowire.
In order to suppress the skyrmion Hall effect, previous works have suggested introducing other degrees of freedom, such as antiferromagnetic interactions~\cite{Zhang2015b,Zhang2016h} and other magnetic layers~\cite{Huang2017,Wu2017a}, which can compensate the skyrmion Hall effect. However, a straight skyrmion motion is believed to be impossible within one ferromagnetic layer, although such simple structures are preferable for applications. Therefore, elimination of the skyrmion Hall effect within one ferromagnetic layer is of significant importance in both scientific and technological aspects.

\begin{figure}
	\includegraphics[width=8.6cm]{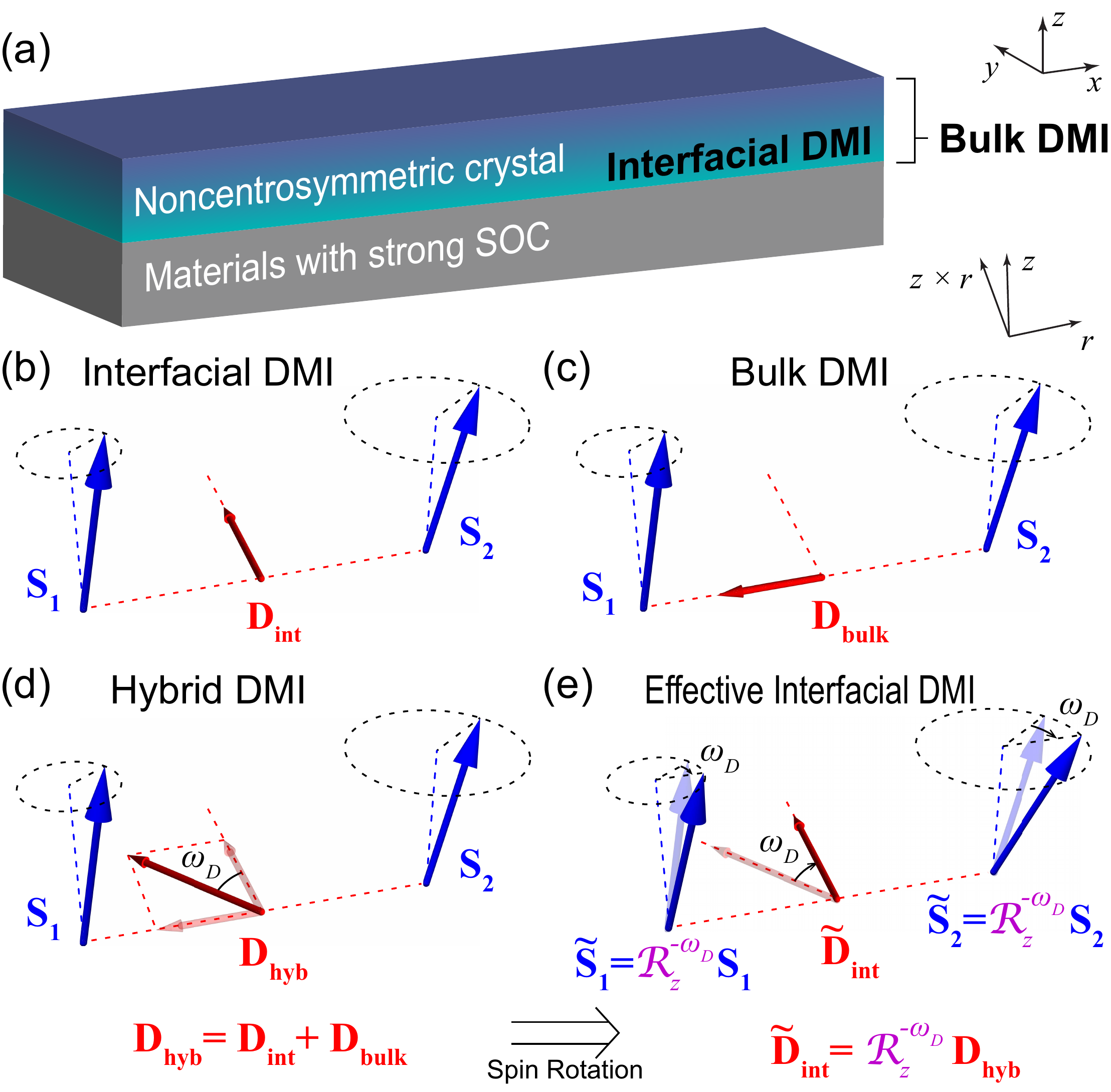}
	\caption{
	(a) Sketch of a system with hybrid DMI, where there exists the bulk DMI as well as the interfacial DMI with the help of strong spin-orbit coupling (SOC).
	Two-spin models for (b) interfacial, (c) bulk, and (d) hybrid DMIs with corresponding DMI vectors (red).
	$\vec{D}_{\rm hyb}$ is tilted by a finite angle $\omega_D$ from the direction of $\vec{D}_{\rm int}$.
	(\textbf{e}) Hybrid DMI is equivalent to the effective interfacial DMI with rotated spins $\tilde{\vec{S}}_i=\mathcal{R}_z^{-\omega_D}\vec{S}_i$, where $\mathcal{R}_z^{\psi}$ denotes the rotation by $\psi$ around the $z$ axis.
		\label{Fig:structure}}
\end{figure}

In this paper, we show how to eliminate the skyrmion Hall effect in a single ferromagnet by utilizing an additional symmetry breaking.
We consider ferromagnets in which both structural inversion symmetry and crystal centrosymmetry are broken~\cite{Rowland2015}, where the DMI arises in the form of a mixture of the interfacial and bulk DMIs. We call this mixture \emph{hybrid} DMI throughout this paper.
The systems with hybrid DMI include crystals with symmetry classes C$_n$~\cite{Cortes-Ortuno2013} and the state-of-the-art multilayer systems consisting of thin chiral magnets (such as thin B20 compounds) and nonmagnetic materials with strong spin-orbit coupling (such as heavy metals like Pt~\cite{Yang2015b} and topological insulators like Bi$_2$Se$_3$~\cite{Han2017}), which is depicted in Fig.~\ref{Fig:structure}(a). We examine the skyrmion motion driven by spin-orbit torque (SOT), whose examples include a spin Hall current injection from heavy metal~\cite{Liu2012,Seo2012,Sinova2015} and the spin-charge conversion from topological surface states~\cite{Mellnik2014}. We find that the skyrmion Hall angle is suppressed for one skyrmion polarity and enhanced for the other.

We explicitly propose experiments on how to control the skyrmion Hall angle by various means. By noting that systems with hybrid DMI can be mapped onto systems with purely interfacial DMI by a global spin rotation, we derive the effective spin torque and current that are acting on hybrid systems. In particular, we find that the direction of the effective current deviates from the applied current direction and depends on the relative strengths of the two DMIs. In total, the skyrmion Hall angle is asymmetric and is tunable by changing the relative DMI strengths. We make several suggestions to eliminate the skyrmion Hall angle for one polarity. As a result, one can achieve a skyrmion motion along the current in a single ferromagnetic layer without the need of interactions with another layer or another sublattice. Our paper suggests a way to exploit symmetry breaking for eliminating detrimental phenomena.

This paper is organized as follows. In Sec.~\ref{Sec:Equilibrium}, we introduce a rotational mapping which maps the hybrid DMI to an effective interfacial DMI. In Sec.~\ref{Sec:Nonequilibrium}, we apply the mapping to a nonequilibrium situation to derive an effective spin torque. In Sec.~\ref{Sec:Asymmetric SkHE}, we show the asymmetry of the skyrmion Hall effect and possibility for making it zero. In Sec.~\ref{Sec:Discussion}, we make more remarks on our theory and suggest various ways to tune the skyrmion Hall angle to zero. In Sec.~\ref{Sec:Summary}, we summarize the paper. Appendices include some useful information which are not directly related to the main flow of our paper.

\section{Hybrid DMI as an effective interfacial DMI\label{Sec:Equilibrium}} We first introduce a mapping of the hybrid DMI to an effective DMI. In contrast to Ref.~\cite{Rowland2015}, where the hybrid DMI is mapped to an effective bulk DMI, in this paper we map it to an interfacial one, in order to directly apply previous knowledge~\cite{Jiang2016,Litzius2017} on the SOT-driven skyrmion motion.
To provide insight, we first present the main idea behind the mapping within a simple two-spin model [Fig.~\ref{Fig:structure}(b)--\ref{Fig:structure}(e)]. The DMI energy for two spins $\vec{S}_1$ and $\vec{S}_2$ that are distant along the $r$ direction takes the form $E_{\rm DMI}=\vec{D}\cdot(\vec{S}_1\times\vec{S}_2)$ for the DMI vector $\vec{D}$~\cite{Dzyaloshinskii1957,Moriya1960b}.
For the interfacial and bulk DMIs [Figs.~\ref{Fig:structure}(b) and \ref{Fig:structure}(c)], the DMI vectors are given by $\vec{D}_{\rm int}=D_{\rm int}\vhat{z}\times\vhat{r}$ and $\vec{D}_{\rm bulk}=-D_{\rm bulk}\vhat{r}$ respectively, where $D_{\rm int/bulk}$ refers to the strength of each DMI, $\vhat{r}$ is the unit vector along $r$, and $\vhat{z}$ is the interface normal direction. The hybrid DMI [Fig.~\ref{Fig:structure}(d)] is described by the hybrid vector $\vec{D}_{\rm hyb}=\vec{D}_{\rm int}+\vec{D}_{\rm bulk}$, which is tilted by an angle $\omega_D$ from $\vec{D}_{\rm int}$, i.e.,\ $D_{\rm int}=D_{\rm hyb}\cos\omega_D$ and $D_{\rm bulk}=D_{\rm hyb}\sin\omega_D$.
From the spin rotation $\mathcal{R}_z^{-\omega_D}$ [Figs.~\ref{Fig:structure}(d) and \ref{Fig:structure}(e)] and the rotational invariance of triple scalar products [Eq.~(\ref{Eq(S):inveriance triple dot})], we obtain the following \emph{effective} interfacial DMI:
  \begin{equation}
\vec{D}_{\rm hyb}\cdot(\vec{S}_1\times\vec{S}_2)=\tilde{\vec{D}}_{\rm int}\cdot(\tilde{\vec{S}}_1\times\tilde{\vec{S}}_2),
\label{Eq:two spin rotation}
\end{equation}
where $\tilde{\vec{D}}_{\rm int}\equiv\mathcal{R}_z^{-\omega_D}\vec{D}_{\rm hyb}=D_{\rm hyb}\vhat{z}\times\vhat{r}$ is the effective interfacial DMI vector for \emph{rotated} spins $\tilde{\vec{S}}_i=\mathcal{R}_z^{-\omega_D}\vec{S}_i$.
Therefore, the physical properties of spins subject to the hybrid DMI are given by those of \emph{rotated} spins under the effective interfacial DMI, provided that all other energy contributions are invariant under $\mathcal{R}_z^{-\omega_D}$. In real crystals, however, the continuous rotational symmetry of energy contributions is not strictly valid. In Appendix~\ref{Sec(A):Discrete symmetry}, we mathematically prove that the main conclusion of our paper is unaltered even for real crystals lacking continuous rotational symmetry.

\begin{figure}
	\includegraphics[width=8.6cm]{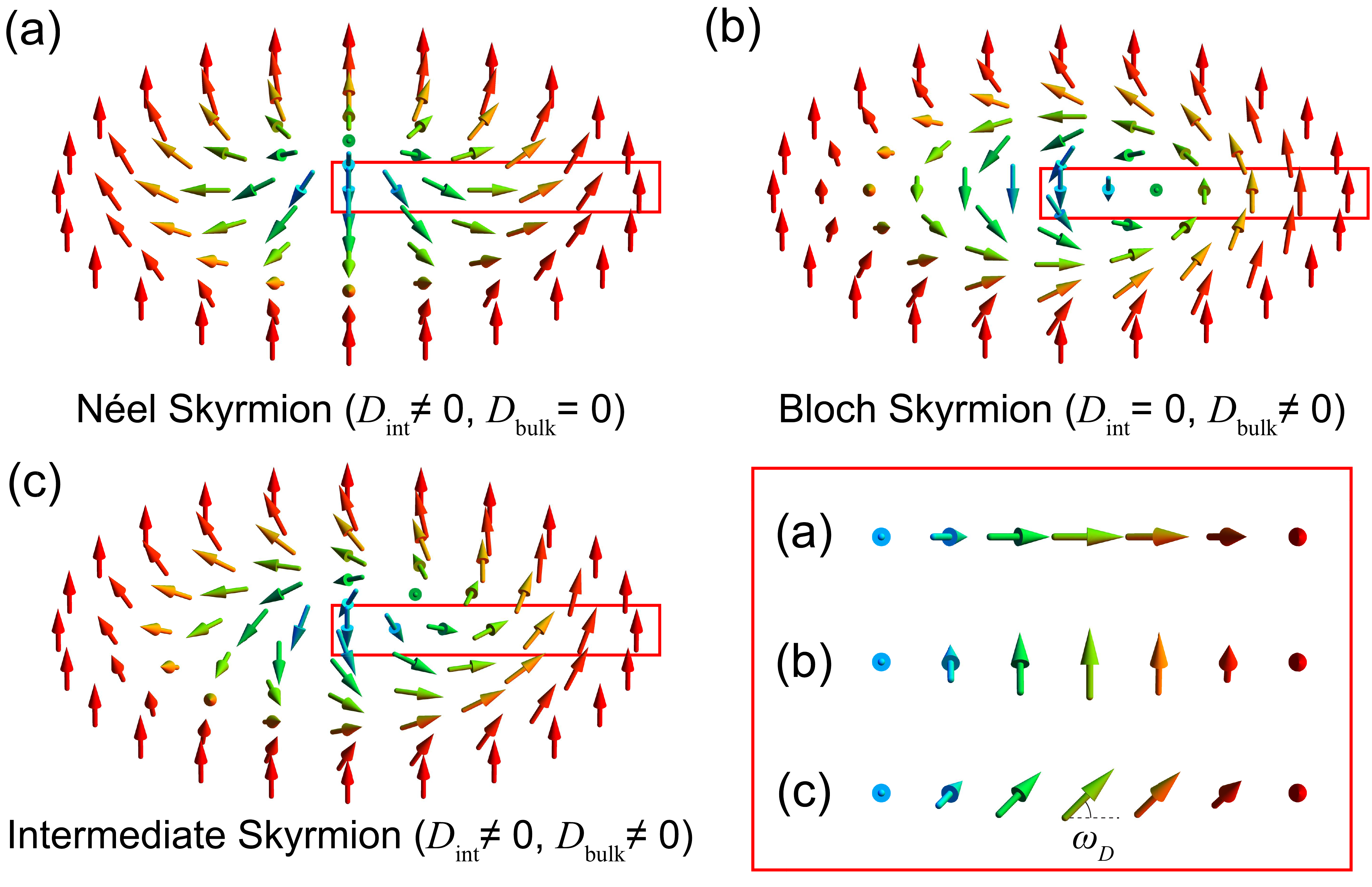}
	\caption{
	Skyrmions stabilized by (a) interfacial DMI, (b) bulk DMI, and (c) hybrid DMI. They can be transformed into each other via a global spin rotation. (inset) Top view of the red-boxed regions, each of which shows (a) N\'{e}el wall, (b) Bloch wall, and (c) an intermediate wall, respectively.\label{Fig:Skyrmions}}
\end{figure}

To apply this insight to continuous magnetic systems, we start from the magnetic energy functional $E=E_0[\vec{m}]+E_{\rm DMI}[\vec{m}]$ for a two-dimensional ferromagnet. The non-DMI contribution is given by $E_0[\vec{m}]=\int d^3r [A(|\partial_x\vec{m}|^2+|\partial_y\vec{m}|^2)-K_zm_z^2]$ where $A$ is the exchange stiffness, and $\vec{m}=\vec{m}(x,y)$ is the unit vector along the local magnetization, and $K_z$ is the uniaxial anisotropy. We consider a system with perpendicular magnetic anisotropy ($K_z>0$), which is preferable for applications~\cite{Jung2008} and naturally arises from the interfacial spin-orbit coupling~\cite{Kim2016f,Chaudhary2018} or the topological surface states~\cite{Kim2017f,Fanchiang2018}.
Regarding dipolar interactions, the DMI typically dominates over them in the properties of magnetic textures~\cite{Iwasaki2013} although they can be important in the absence of DMI~\cite{Zhang2017h}. In Sec.~\ref{Sec:Simulation}, we present the results of micromagnetic simulations, which confirms that dipolar interactions do not change our main conclusion qualitatively (see Fig.~\ref{Fig:Simulation} for more information). $E_{\rm DMI}[\vec{m}]$ is the hybrid DMI contribution,
\begin{subequations}
\begin{equation}
E_{\rm DMI}[\vec{m}]=D_{\rm int}\mathcal{E}_{\rm int}[\vec{m}]+D_{\rm bulk}\mathcal{E}_{\rm bulk}[\vec{m}],\label{Eq:total DMI}
\end{equation}
where $D_{\rm int}$ and $D_{\rm bulk}$ are the strengths of the interfacial and bulk DMI respectively, and
the continuous forms of the DMIs are given by $\mathcal{E}_{\rm int}=\int d^3r \sum_{u=x,y}(\vhat{z}\times\vhat{u})\cdot(\vec{m}\times\partial_u\vec{m})$, and $\mathcal{E}_{\rm bulk}=\int d^3r~\vec{m}\cdot(\nabla\times\vec{m})$.
The continuous analog of Eq.~\eqref{Eq:two spin rotation} gives
\begin{equation}
E=E_0[\tilde{\vec{m}}]+D_{\rm hyb}\mathcal{E}_{\rm int}[\tilde{\vec{m}}],\label{Eq:energy functional mu}
\end{equation}
\end{subequations}
with the \emph{globally} rotated magnetization $\tilde{\vec{m}}=\mathcal{R}_z^{-\omega_D}\vec{m}$. Note that the effective DMI strength $|\tilde{\vec{D}}_{\rm int}|$ is identical to $D_{\rm hyb}$. This effective model for $\tilde{\vec{m}}$ contains the interfacial DMI only and thus stabilizes N\'{e}el skyrmions [Fig.~\ref{Fig:Skyrmions}(a)].
Then, one can directly apply previously developed knowledge on N\'{e}el skyrmions~\cite{Jiang2016,Litzius2017} to examine physical properties of $\vec{m}$ by applying the inverse rotation. For example, hybrid DMI stabilizes a structure interpolating between N\'{e}el and Bloch skyrmions (Fig.~\ref{Fig:Skyrmions})~\cite{Rowland2015}, which we call an \emph{intermediate skyrmion}. The stability of intermediate skyrmions are immediately given by previous studies for N\'{e}el and Bloch skyrmions~\cite{Leonov2016a,Buettner2018} if the DMI strength is replaced by $D_{\rm hyb}$. Note that the stability and the equilibrium properties (such as the skyrmion diameter) are independent of $\omega_D$. Our micromagnetic simulations show that this is a good approximation even in the presence of dipolar interactions [see Fig.~\ref{Fig:Simulation}(a) in Sec.~\ref{Sec:Simulation} for more information].

\section{Nonequilibrium: Effective spin torque\label{Sec:Nonequilibrium}} To describe the dynamics of intermediate skyrmions in nonequilibrium, we start from the Landau-Lifshitz-Gilbert equation:
\begin{subequations}
\begin{equation}
\partial_t\vec{m}=-\gamma\vec{m}\times\vec{H}_{\rm eff}+\alpha\vec{m}\times\partial_t\vec{m}+\vec{T}[\vec{m}],\label{Eq:LLG m}
\end{equation}
where $\vec{H}_{\rm eff}=M_s^{-1}(2A\nabla^2\vec{m}+2K_zm_z\vhat{z}+D_{\rm int}h_{\rm int}[\vec{m}]+D_{\rm bulk}h_{\rm bulk}[\vec{m}])$ is the effective magnetic field for $\vec{m}$,  $h_{\rm int/bulk}=-\delta \mathcal{E}_{\rm int/bulk}/\delta\vec{m}$,  $\gamma$ is the gyromagnetic ratio, $\alpha$ is the Gilbert damping parameter, and $\vec{T}[\vec{m}]$ refers to the spin torque induced by an applied current.
In terms of $\tilde{\vec{m}}$, Eq.~\eqref{Eq:LLG m} is equivalent to
\begin{equation}
\partial_t\tilde{\vec{m}}=-\gamma\tilde{\vec{m}}\times\tilde{\vec{H}}_{\rm eff}+\alpha\tilde{\vec{m}}\times\partial_t\tilde{\vec{m}}+\tilde{\vec{T}}[\tilde{\vec{m}}],\label{Eq:LLG mu}
\end{equation}
where $\tilde{\vec{H}}_{\rm eff}=M_s^{-1}(2A\nabla^2\tilde{\vec{m}}+2K_z\tilde{m}_{z}\vhat{z}+D_{\rm hyb}h_{\rm int}[\tilde{\vec{m}}])$ is the effective field for $\tilde{\vec{m}}$ and contains an (effective) interfacial DMI contribution \emph{only}. $\tilde{\vec{T}}[\tilde{\vec{m}}]=\mathcal{R}_z^{-\omega_D}\vec{T}[\vec{m}]$ is the \emph{effective} spin torque acting on $\tilde{\vec{m}}$, which is discussed extensively below.
Equation~(\ref{Eq:LLG mu}) implies that the dynamics of an intermediate skyrmion is described by that of a N\'{e}el skyrmion with an effective spin torque (see Table~\ref{Tab:summary}).

\begin{table}[b]
	\renewcommand{\arraystretch}{1.5}
	\begin{tabular}{c|c|c|c|c|c}
		Frame&Spins&DMI&Skyrmion&Spin torque&Current\\\hline\hline
		Lab frame&$\vec{m}$&Hybrid&Intermediate&$\vec{T}[\vec{m}]$&$\vec{j}$\\\hline
		Rotated&$\tilde{\vec{m}}$&Interfacial&N\'{e}el&$\tilde{\vec{T}}[\tilde{\vec{m}}]$&$\tilde{\vec{j}}$\\
	\end{tabular}
	\caption{\label{Tab:summary} Analogous descriptions of intermediate skyrmions (upper row) and N\'{e}el skyrmions (lower row). Previously developed knowledge on the N\'{e}el skyrmion dynamics gives the intermediate skyrmion dynamics without explicit calculations.}
\end{table}

There are three remarks. First, our theory does not require any ansatz on the internal structures of skyrmions, thus it is valid even when a current-induced deformation of skyrmion~\cite{Litzius2017} and higher order  skyrmions~\cite{Rozsa2017} are taken into account. Second, our mapping still works for more general models including a generalized form of the interfacial DMI~\cite{Vida2016,Rozsa2017}. Third, although introduction of in-plane fields or anisotropies breaks the rotational symmetry, our theory is still valid if they appear as rotated terms in $\tilde{\vec{H}}_{\rm eff}$.

The effective spin torque in terms of $\tilde{\vec{m}}$ is
\begin{equation}
\tilde{\vec{T}}[\tilde{\vec{m}}]=R_z^{-\omega_D}\vec{T}[R_z^{\omega_D}\tilde{\vec{m}}].\label{Eq:effective spin torque}
\end{equation}
\end{subequations}
Equation~(\ref{Eq:LLG mu}) with Eq.~(\ref{Eq:effective spin torque}) is the first central result of this paper.
It implies that spin torques that are covariant under rotation, i.e.,  $\vec{T}[\mathcal{R}_z^{\omega_D}\tilde{\vec{m}}]=\mathcal{R}_z^{\omega_D}\vec{T}[\tilde{\vec{m}}]$, impose the same dynamics for the intermediate skyrmion as for a N\'eel skyrmion, since $\tilde{\vec{T}}[\tilde{\vec{m}}]=\vec{T}[\tilde{\vec{m}}]$.
This is in particular the case for adiabatic~\cite{Tatara2004} and nonadiabatic~\cite{Zhang2004} spin-transfer torques as explicitly shown in Appendix~\ref{Sec(A):conventional spin torque}.

\begin{figure}
	\includegraphics[width=8.6cm]{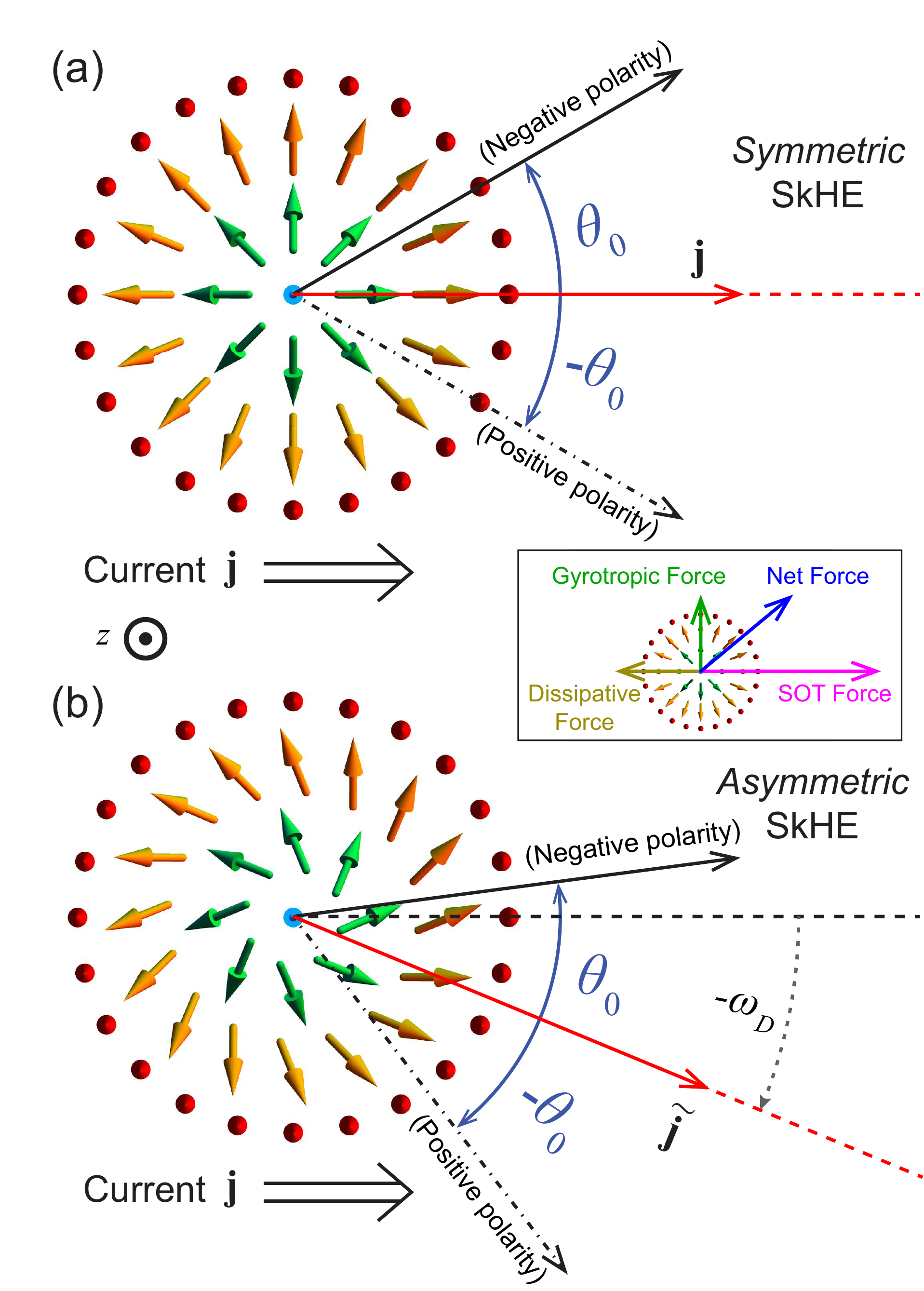}
	\caption{Top-viewed sketches of SOT-induced motions of a N\'{e}el skyrmion and an intermediate skyrmion and their skyrmion Hall effect (SkHE). (a) For the N\'{e}el skyrmion, the skyrmion Hall effect is symmetric for each skyrmion polarity (black solid and black dot-dashed arrows). (b) An intermediate skyrmion feels an effective current $\tilde{\vec{j}}$ which is rotated by $-\omega_D$. Thus, we obtain an asymmetric skyrmion Hall effect [Eq.~(\ref{Eq:asymmetric SkHE})]. (inset) Forces acting on the N\'{e}el skyrmion at the initial state.
		\label{Fig:asymmetric SkHE}}
\end{figure}

On the other hand, when the spin torque includes a term not covariant under rotation, the intermediate skyrmion shows a \emph{qualitatively} different dynamics from a N\'{e}el skyrmion.
Crucially, we show below that SOTs do lead to qualitatively different dynamics for intermediate skyrmions. SOTs arise in systems composed of materials with strong spin-orbit coupling and magnetic layers as we consider in this work. Moreover, they dominate the adiabatic and nonadiabatic spin torques~\cite{Liu2012,Mellnik2014,Emori2013,Ryu2013}.
Examples of such SOTs include a spin Hall current injection from heavy metal~\cite{Liu2012,Seo2012,Sinova2015} and the spin-charge conversion from topological surface states~\cite{Mellnik2014}. They take the form
\begin{subequations}\label{Eq:SOT a-b}
\begin{equation}
\vec{T}[\vec{m}]=\tau_{\rm f}\vec{m}\times(\vhat{z}\times\vec{j})+ \tau_{\rm d}\vec{m}\times[\vec{m}\times(\vhat{z}\times\vec{j})]\label{Eq:SOT}
\end{equation}
where $\tau_{\rm f}$ and $\tau_{\rm d}$ are coefficients for the field-like SOT~\cite{Manchon2008,Matos-Abiague2009c} and the damping-like SOT~\cite{Wang2012b,Kim2012b,Pesin2012,Kurebayashi2014}, respectively. They are proportional to the spin Hall angle of the heavy metal or the spin-charge conversion efficiency of the topological surface states.
The roles of internally generated SOT in the ferromagnet~\cite{Lee2017}, which is another type of SOT, are discussed in Appendix~\ref{Sec(A):Internal spin torque}. Applying Eq.~(\ref{Eq:effective spin torque}) to Eq.~\eqref{Eq:SOT} gives the effective SOT
\begin{equation}
\tilde{\vec{T}}[\tilde{\vec{m}}]=\tau_{\rm f}\tilde{\vec{m}}\times(\vhat{z}\times\tilde{\vec{j}})+ \tau_{\rm d}\tilde{\vec{m}}\times[\tilde{\vec{m}}\times(\vhat{z}\times\tilde{\vec{j}})].\label{Eq:effective SOT}
\end{equation}
\end{subequations}
Equation~(\ref{Eq:effective SOT}) is straightforwardly verified by using the rotational covariance of cross products [Eq.~(\ref{eq:cross})] as well as by a direct calculation. Despite the resemblance to Eq.~\eqref{Eq:SOT}, the \emph{effective} current $\tilde{\vec{j}}=R_z^{-\omega_D}\vec{j}$, felt by the rotated magnetization $\tilde{\vec{m}}$, enters in Eq.~(\ref{Eq:effective SOT}) (see Fig.~\ref{Fig:asymmetric SkHE}), thus $\tilde{\vec{T}}[\tilde{\vec{m}}]\neq\vec{T}[\tilde{\vec{m}}]$.
The reason for obtaining a different torque is the appearance of the fixed direction $\vhat{z}\times\vec{j}$ independent of $\tilde{\vec{m}}$, which breaks the rotational covariance of the SOT.
Consequently, the dynamics of intermediate skyrmions is given by that of N\'{e}el skyrmions under the effective current $\tilde{\vec{j}}$, which has a \emph{different} direction from the applied current $\vec{j}$. This results in significant implications on current-induced dynamics of intermediate skyrmions as we demonstrate in Sec.~\ref{Sec:Asymmetric SkHE}.

\section{Asymmetric Skyrmion Hall effect and zero Hall angle\label{Sec:Asymmetric SkHE}} Before describing the dynamics of intermediate skyrmions, we first briefly review the SOT-induced dynamics of N\'{e}el skyrmions~\cite{Jiang2016,Litzius2017}. In Fig.~\ref{Fig:asymmetric SkHE}(a), we sketch a N\'{e}el skyrmion with the negative polarity (i.e., $m_z=-1$ at the core). There are several forces acting on it (inset). As a result of SOT, a force along the current $\vec{j}$ is exerted on the N\'{e}el skyrmion, inducing its motion. While the skyrmion is moving, there arises a dissipative (gyrotropic) force along (perpendicular to) its velocity such that the steady state skyrmion motion direction is deviated from $\vec{j}$ with a nonzero tilting angle $\theta_0$ (black solid arrow). This effect is called the skyrmion Hall effect. For the positive polarity (the core magnetization $m_z=1$), it is subject to an opposite gyrotropic force, thus the transverse velocity is opposite (black dot-dashed arrow). Hence, the skyrmion Hall angles for each polarity of the N\'{e}el skyrmion have the same magnitudes but opposite signs: $\theta_{\rm SkHE,\pm}^{\rm N}=\mp \theta_0$, where the signs refer to the polarities and the superscript N refers to N\'{e}el skyrmions [Fig.~\ref{Fig:asymmetric SkHE}(a)].

\begin{figure*}
	\includegraphics[width=17.2cm]{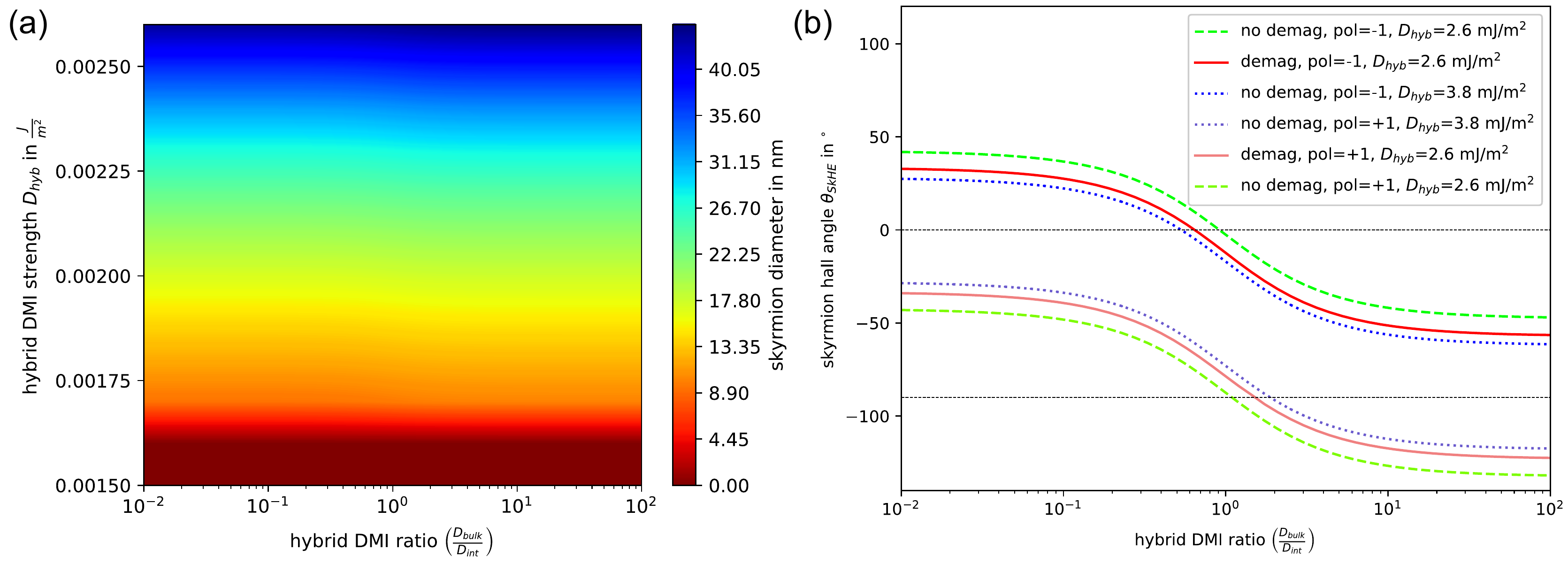}
	\caption{(a) Skyrmion diameters for various $D_{\rm bulk}/D_{\rm int}$ ratios including dipolar interactions. (b) Skyrmion Hall angles for various $D_{\rm bulk}/D_{\rm int}$ ratios and skyrmion polarities, with and without dipolar interactions (denoted by demag and no demag, respectively).
		\label{Fig:Simulation}}
	\end{figure*}

Applying Table~\ref{Tab:summary} allows a direct extension of the above knowledge to the intermediate skyrmion dynamics. In the rotated frame, the magnetic texture forms a N\'{e}el skyrmion which is subject to the effective current $\tilde{\vec{j}}$. Therefore, the nonzero tilting angle $\pm\theta_0$ arises symmetrically around $\tilde{\vec{j}}$, thus is \emph{asymmetric} around $\vec{j}$. Getting back to the laboratory frame immediately gives the intermediate skyrmion dynamics in Fig.~\ref{Fig:asymmetric SkHE}(b) and its Hall angle
\begin{equation}
\theta_{\rm SkHE,\pm}^{\rm I}=-\omega_D\mp\theta_0,\label{Eq:asymmetric SkHE}
\end{equation}
which is the second central result of this paper. Here the superscript I refers to intermediate skyrmions. Equation~(\ref{Eq:asymmetric SkHE}) is also confirmed by the collective coordinate approach~\cite{Thiele1973} (Appendix~\ref{Sec(A):Thiele}) and by micromagnetic simulations (Sec.~\ref{Sec:Simulation}). Depending on the skyrmion polarity, the magnitude of the skyrmion Hall angle is enhanced or suppressed. For device applications, the skyrmion polarity with the lower magnitude can be chosen via changing the background magnetization.

\begin{table}[b]
\begin{tabular}{c|c|c|c|c|c}		Compound&MnSi&MnGe&FeGe&Mn$_{1-x}$Fe$_x$Ge&Mn$_{1-x}$Ir$_x$Si\\\hline\hline
			$|D_{\rm bulk}|$ &0.81&0.54&0.87&0.00--1.95&0.76--0.81
		\end{tabular}
		\caption{\label{Tab:Bulk DMI} Reported strengths of the bulk DMI ($\mathrm{mJ/m^2}$) for various B20 compounds. The strengths for Mn$_{1-x}$Fe$_x$Ge are taken from Refs.~\cite{Gayles2015,Turgut2018} and those for the other are taken from Ref.~\cite{Dhital2017} and references therein. For conversions between strengths for discrete models to continuum models, we use the lattice constants in Ref.~\cite{Wilhelm} and references therein.
		}
\end{table}

Equation~(\ref{Eq:asymmetric SkHE}) implies the possibility to \emph{completely} eliminate the skyrmion Hall angle for one polarity when $|\theta_0|=|\omega_D|$. From $\tan\omega_D=D_{\rm bulk}/D_{\rm int}$, the relative strength of the DMIs is crucial to realize this. For an estimation for this possibility, we use the experimentally observed skyrmion Hall angle $|\theta_0|\approx 30^\circ$~\cite{Jiang2016,Litzius2017} and common bulk DMI strengths range $|D_{\rm bulk}|=(0.0~\mathrm{to}~2.0)~\mathrm{mJ/m^2}$ (Table~\ref{Tab:Bulk DMI}).
To achieve $|\theta_0|=|\omega_D|$, this corresponds to a range of interfacial DMI strengths of $|D_{\rm int}|=(0.0~\mathrm{to}~3.4)~\mathrm{mJ/m^2}$, which coincides well with the typical range of the interfacial DMI strength, $|D_{\rm int}|=(0.0~\mathrm{to}~3.0)~\mathrm{mJ/m^2}$, reported for ferromagnets in contact with Pt, Ta, and Ir~\cite{Emori2014,Heinonen2016,Moreau-Luchaire2016,Pizzini2014,Belmeguenai2015,Yang2015b}. In Sec.~\ref{Sec:Discussion}, we propose various ways to realize the zero Hall angle experimentally.

\section{Micromagnetic simulations\label{Sec:Simulation}}

We perform micromagnetic simulations to verify our assumptions and reproduce the key results obtained from our simple analytic formalism in the more realistic case including dipolar interactions. Our simulations were performed with MuMax3~\cite{Vansteenkiste2014} available at \url{http://mumax.github.io/}
to solve the Landau-Lifshitz-Gilbert equation. The code is modified to simulate both interfacial and bulk DMIs simultaneously. We simulate a system containing $128\times128\times1$ cells of $1~\mathrm{nm}$ cell size and $0.5~\mathrm{nm}$ thickness and impose a periodic boundary condition. The following material parameters were used: the exchange stiffness $A=10^{-11}~\mathrm{J/m}$, the saturation magnetization $M_s=6\times10^5~\mathrm{A/m}$, the perpendicular magnetic anisotropy $K_z=5\times10^5~\mathrm{J/m^3}$, the Gilbert damping $\alpha=1$, and various values of the DMI strengths.

First, we examine the static properties of intermediate skyrmions. Our simulations show that the stability and the size are basically the same for various $D_{\rm bulk}/D_{\rm int}$ ratios when $D_{\rm hyb}$ is fixed [Fig.~\ref{Fig:Simulation}(a)]. The slight deviations from exact horizontal lines originate from dipolar interactions, which is confirmed by additional simulations without dipolar interactions (not shown). This is in perfect agreement with our analytic theory.

	We then perform the simulation for the skyrmion Hall angle as presented in Fig.~\ref{Fig:Simulation}(b). The simulation shows that one can always find a proper $\omega_D$ value which makes the skyrmion Hall angle zero. Thus, the main conclusion of our work is unchanged.

	More quantitatively, for the parameters chosen, the zero Hall angle occurs at $D_{\rm bulk}/D_{\rm int}=0.91$ without dipolar interactions (green line) and $D_{\rm bulk}/D_{\rm int}=0.64$ with dipolar interactions for $D_{\rm hyb}=2.6~\mathrm{mJ/m^2}$ (red line). The nonnegligible difference between the two values originates from the different skyrmion diameter. $\theta_0$ is dependent on the skyrmion diameter~\cite{Jiang2016,Litzius2017} and we also check the dependency in our simulations (not shown). To match the skyrmion diameters, we perform another simulation with $D_{\rm hyb}=3.8~\mathrm{mJ/m^2}$ (blue line) where the skyrmion diameter is approximately given by that of the red line. The zero Hall angle occurs at $D_{\rm bulk}/D_{\rm int}=0.53$, which deviates the value for the red line by a small difference. This difference could originate from different domain wall widths, which is also dependent on the DMI value and is proportional to $\tan\theta_0$~\cite{Litzius2017}.

\section{Discussion\label{Sec:Discussion}}

There are three remarks. First, it is interesting that the skyrmion Hall effect is eliminated while the topological Hall effect of electrons is still present. Second, the zero skyrmion Hall angle is preserved under current reversal, which is apparent by rotation of the sample around the $z$ axis by $\pi$. Third, $\theta_0$ can have linear current dependence due to disorder~\cite{Jiang2016,Reichhardt2015} and higher order distortions~\cite{Litzius2017} particularly in a low current density regime. In this regime, the zero Hall angle occurs only at a particular current density. However, our primary interest is a more high current density regime where the skyrmion information is delivered with a high speed. In this regime, the skyrmion Hall angle saturates, thus the zero Hall angle is stable against current variations. The saturation current density is on the order of $10^{11}~\mathrm{A/m^2}$~\cite{Jiang2016}, which is the typical order of magnitude used for racetrack applications. The numerous ongoing researches on the spin-orbit torque efficiency will even reduce it in the future. These justify the regime of our consideration.

In order to realize the zero Hall angle, tuning the DMI strengths is crucial for achieving $|\theta_0|=|\omega_D|$. We present various ways to tune the DMI strength in experiment. (i) Thickness variations:
Since $D_{\rm int}$ is an interface contribution, varying thickness of the ferromagnetic layer~\cite{Yang2015b,Chaudhary2018} or the adjacent nonmagnetic layer~\cite{Cao2017,Chaudhary2018} changes $D_{\rm int}$ significantly. (ii) Modifying interface properties: Controlling interface properties by oxygen~\cite{Belabbes2015} or $\mathrm{Ar}^+$ irradiation with various irradiation energies~\cite{Balk2017} allows a fine tuning of $D_{\rm int}$ and even changing its sign. Moreover, insertion of a layer is also possible to change $D_{\rm int}$ significantly~\cite{Hrabec2014,Vida2016}. (iii) Voltage-induced control of interfacial DMI: A perpendicular gate voltage changes the asymmetry along the $z$ direction, thus it changes the interfacial DMI strength~\cite{Nawaoka2015,Yang2016d}, \emph{dynamically}. (iv) Composition variation: As demonstrated in Mn$_{1-x}$Fe$_x$Ge for various $x$~\cite{Gayles2015,Shibata2013,Grigoriev2013,Turgut2018,Koretsune2015}, varying the composition of the bulk material allows changing the magnitude and sign of $D_{\rm bulk}$. (v) Strain: $D_{\rm bulk}$ can also be controlled dynamically by using a strain~\cite{Shibata2015,Chacon2015,Rowland2015}. Since the change of the bulk DMI is anisotropic, more quantitative studies are required for this direction.

\section{Summary\label{Sec:Summary}}

In summary, we demonstrate an asymmetric skyrmion Hall effect driven by SOTs in the presence of mixture of the interfacial and bulk DMIs, where an intermediate skyrmion interpolating between a N\'{e}el and a Bloch skyrmions
is stabilized. We develop an effective theory for the dynamics of intermediate skyrmions by mapping it onto that of effective N\'eel skyrmions. This allows examining the intermediate skyrmion dynamics without additional explicit calculations. The effective current felt by the skyrmion is tilted by an angle determined by the relative strength of the two DMIs. As a result, the skyrmion Hall angle is asymmetric for each skyrmion polarity. By an explicit estimation with realistic values, we demonstrate the possibility to completely eliminate the skyrmion Hall effect for one polarity even without the help of antiferromagnetic interactions or another layer. We make several suggestions for the experimental realization of it, via tuning the strengths of the DMIs.
Our work provides a way to exploit symmetry breaking for removing detrimental phenomena as well as advances next generation spintronic applications based on skyrmions.

\begin{acknowledgments}
We acknowledge A.~Thiaville, M.~Kl\"{a}ui, A.~Bogdanov, J.~Sinova, K.~Litzius, and K.~Hals for fruitful discussions. This work is supported by the German Research Foundation (DFG) (No. EV 196/2-1 and No. SI 1720/2-1), the Alexander von Humboldt Foundation, the Transregional Collaborative Research Center (SFB/TRR) 173 SPIN+X. K.-W.M was supported by Creative Materials Discovery Program through the National Research Foundation of Korea (NRF) funded by the Ministry of Science, ICT and Future Planning (2015M3D1A1070467) and by the National Research Council of Science \& Technology (NST) grant (No. CAP-16-01-KIST) by the Korea government (MSIP). N.K. gratefully acknowledges financial support of the German Academic Scholarship Foundation and support by the Graduate School of Excellence Materials Science in Mainz (MAINZ).
\end{acknowledgments}

\begin{appendix}

\section{Rotational transformation of vector products\label{Sec(A):Rotational transform}}
	\subsection{Rotational invariance of triple scalar products}
	Given arbitrary three-dimensional vectors $\vec{a}$, $\vec{b}$, and $\vec{c}$ and an arbitrary rotation matrix $\mathcal{R}$, the triple scalar product $\vec{c}\cdot(\vec{a}\times\vec{b})$ is invariant under $\mathcal{R}$, i.e.,
	\begin{equation}	\vec{c}\cdot(\vec{a}\times\vec{b})=(\mathcal{R}\vec{c})\cdot[(\mathcal{R}\vec{a})\times(\mathcal{R}\vec{b})].
	\label{Eq(S):inveriance triple dot}
	\end{equation}
	This can be intuitively understood by the fact that the triple scalar product is nothing but the volume of the parallelepiped whose edges are given by $\vec{a}$, $\vec{b}$, and $\vec{c}$, which does not change under any rotation.

	\subsection{Rotational covariance of cross products}
	Given arbitrary three-dimensional vectors $\vec{a}$ and $\vec{b}$ and an arbitrary rotation matrix $\mathcal{R}$, the cross product $\vec{a}\times\vec{b}$ is rotationally covariant, i.e.,
	\begin{equation} (\mathcal{R}\vec{a})\times(\mathcal{R}\vec{b})=\mathcal{R}(\vec{a}\times\vec{b}).
	\label{eq:cross}
	\end{equation}
	This means that the rotation of a cross product is nothing but the cross product of the rotated vectors, which can be intuitively understood by the right-hand rule for cross products.

	\subsection{Rotational covariance of the conventional spin-transfer torques\label{Sec(A):conventional spin torque}}
	We start from the conventional spin-transfer torques in the form of adiabatic~\cite{Tatara2004} and nonadiabtic~\cite{Zhang2004} torques $ \vec{T}_{\rm conv}(\vec{m})=b(\vec{j}\cdot\nabla)\vec{m}+c\vec{m}\times(\vec{j}\cdot\nabla)\vec{m}$ and show
	\begin{equation}
	\vec{T}_{\rm conv}(\mathcal{R}\vec{m})=\mathcal{R}\vec{T}_{\rm conv}(\vec{m}),
	\end{equation} for an arbitrary global rotation $\mathcal{R}$. Here $b$ and $c$ are coefficients for adiabatic and nonadiabatic spin-transfer torques and $\vec{j}$ is the applied current.
	
	Since $\mathcal{R}$ is a global rotation which is independent of position, it commutes with gradients. This means that the adiabatic spin-transfer torque is covariant under $\mathcal{R}$: $b(\vec{j}\cdot\nabla)(\mathcal{R}\vec{m})=\mathcal{R}[b(\vec{j}\cdot\nabla)\vec{m}]$. Additional use of the rotational covariance of the cross product [Eq.~\eqref{eq:cross}] leads us to that the nonadiabatic spin-transfer torque is also covariant under $\mathcal{R}$: $c(\mathcal{R}\vec{m})\times(\vec{j}\cdot\nabla)(\mathcal{R}\vec{m})=c(\mathcal{R}\vec{m})\times\mathcal{R}[(\vec{j}\cdot\nabla)\vec{m}]=\mathcal{R}[c\vec{m}\times(\vec{j}\cdot\nabla)\vec{m}]$. These two observations show the rotational covariance of the conventional spin-transfer torques $\vec{T}_{\rm conv}(\mathcal{R}\vec{m})=\mathcal{R}\vec{T}_{\rm conv}(\vec{m})$, not resulting in the asymmetric skyrmion Hall effect.

	\section{Breakdown of continuous rotational symmetry\label{Sec(A):Discrete symmetry}}
	We prove here that the main conclusion made in our paper is unchanged for real crystals where continuous rotational symmetry is broken. More precisely, even if continuous rotational symmetry is broken, the discrete rotational symmetry of crystals allows for zero skyrmion Hall angle for proper choice of $\omega_D$.
	
	Consider a crystal with an $n$-fold discrete rotational symmetry around the $z$ axis and denote the skyrmion Hall angle as a function of $\omega_D$ by $\theta_{\mathrm{SkHE},p}(\omega_D)$, where $p$ is the polarity of the skyrmion. Since the system lacks the continuous rotational symmetry, $\theta_{\mathrm{SkHE},p}(\omega_D)$ is not strictly given by $-\omega_D-p\theta_0$ as derived in the main text [Eq.~(\ref{Eq:asymmetric SkHE})]. Nevertheless, we can show that it is always possible to tune the skyrmion Hall angle to zero by the following proof. Since the crystal has an $n$-fold discrete rotational symmetry, all energy terms are invariant under discrete rotation by $\pm 2k\pi/n$, where $k$ is a positive integer smaller than $n$. Then, Eq.~(\ref{Eq:asymmetric SkHE}) is valid for these special angles. We now consider $k$ such that $2k\pi/n>\theta_0$. This choice is always possible since the observed $\theta_0$ does not exceed $\pi/2$. Applying Eq.~(\ref{Eq:asymmetric SkHE}) for the angles $\pm2k\pi/n$ leads to $\theta_{\mathrm{SkHE},p}(2k\pi/n)>0$ and $\theta_{\mathrm{SkHE},p}(2k\pi/n)<0$. Now the intermediate value theorem for continuous functions implies that it is always possible to find an angle $\omega_D$ in the range $-2k\pi/n < \omega_D < 2k\pi/n$ such that $\theta_{\mathrm{SkHE},p}(\omega_D)=0$. This proves that it is always possible to tune the skyrmion Hall angle to zero even when continuous rotational symmetry is broken.

	\section{Roles of internally generated SOT\label{Sec(A):Internal spin torque}}
	
	In the present paper, we consider SOTs originating from the adjacent layer, which do not change their form when the form of the DMI is altered. In contrast, internally generated SOTs within the ferromagnet can change their form when hybrid DMI is taken into account, if they have the same origin as DMI~\cite{Kim2013b,Yin2016}. A previous work~\cite{Lee2017} examines the skyrmion motion driven by the internal SOTs for various types of spin-orbit coupling, including Rashba, Dresselhaus, Weyl, and their mixtures. Due to the correlation between the form of the internal SOTs and DMI, an asymmetric skyrmion Hall effect was not observed in that work. We emphasize that (i) the additional consideration of the internal SOTs does not change the main result of our work, i.e., the asymmetric skyrmion Hall effect. (ii) High resistivities of chiral magnets based on Mn, Si, and Ge~\cite{Stenstrom1972,Petrova2006} (several times larger than Pt for example) imply that the effects of internal SOTs are typically smaller than those of external SOTs. Thus, we claim that the asymmetric skyrmion motion will be the dominant effect.
	
	\section{Collective coordinate approach for the asymmetric skyrmion Hall effect\label{Sec(A):Thiele}}
	We derive the skyrmion Hall effect by using the Thiele approach~\cite{Thiele1973}. This approach is restricted for the rigid regime, where the roles of skyrmion deformation are negligible, but allows calculating the skyrmion Hall angle directly without rotational transformation. We start from the following Landau-Lifshitz-Gilbert equation.
	\begin{equation}
	\partial_t\vec{m}=-\gamma\vec{m}\times\vec{H}_{\rm eff}+\alpha\vec{m}\times\partial_t\vec{m}+\vec{T}(\vec{m}),\label{Eq(S):LLG}
	\end{equation}
	where SOT is given by $\vec{T}(\vec{m})=\tau_{\rm f}\vec{m}\times(\vhat{z}\times\vec{j})+ \tau_{\rm d}\vec{m}\times[\vec{m}\times(\vhat{z}\times\vec{j})]$. Assuming a rigid motion of the skyrmion, let $\vec{m}_0(\vec{r})$ be the skyrmion stabilized by $\vec{m}\times\vec{H}_{\rm eff}|_{\vec{m}=\vec{m}_0}=0$ and its center is chosen by $\vec{r}=0$ without loss of generality. Then the intermediate skyrmion is given by $\vec{m}_0(\vec{r})=(\sin\theta(\rho)\cos(\phi+\omega_D),\sin\theta(\rho)\sin(\phi+\omega_D),\cos\theta(\rho))$, where $\rho=\sqrt{x^2+y^2}$ and $\phi=\arg(x+iy)$, as demonstrated in the main text. $\theta(\rho)$ satisfies $\lim_{\rho\to0}\cos\theta(\rho)=\pm 1$ and $\lim_{\rho\to\infty}\cos\theta(\rho)=\mp 1$ depending on the skyrmion polarity.
	The detailed form of $\theta(\rho)$~\cite{Leonov2016} is not necessary for our calculation.
	
	We assume that, in the presence of SOT, the magnetism profile is $\vec{m}(\vec{r})=\vec{m}_0(\vec{r}-\vec{R}(t))$, where $\vec{R}(t)=(X(t),Y(t))$ is the position of the skyrmion. Then, the Thiele equations are given by $\int [\mathrm{Eq.~(\ref{Eq(S):LLG})}]\cdot(\vec{m}\times\partial_X\vec{m})d^2r=0$ and $\int [\mathrm{Eq.~(\ref{Eq(S):LLG})}]\cdot(\vec{m}\times\partial_Y\vec{m})d^2r=0$. After some algebra, we obtain the following equation of motion for $X(t)$ and $Y(t)$.
	\begin{align}
	\begin{pmatrix}
	0&-G\\G&0
	\end{pmatrix}\begin{pmatrix}
	\dot{X}\\\dot{Y}
	\end{pmatrix}&=\begin{pmatrix}
	\mathcal{D}&0\\0&\mathcal{D}
	\end{pmatrix}\begin{pmatrix}
	\dot{X}\\\dot{Y}
	\end{pmatrix}\nonumber\\
&\quad+\begin{pmatrix}
	\cos\omega_D&\sin\omega_D\\-\sin\omega_D&\cos\omega_D
	\end{pmatrix}\begin{pmatrix}
	F_x^{\rm SOT}\\F_y^{\rm SOT}
	\end{pmatrix},\label{Eq(S):Thiele}
	\end{align}
	where $\dot{X}=X'(t)$, $\dot{Y}=Y'(t)$, $G=-\int\vec{m}_0\cdot(\partial_x\vec{m}_0\times\partial_y\vec{m}_0)d^2r$ is the skyrmion number multiplied by $-4\pi$, $\mathcal{D}=\alpha\int \partial_x\vec{m}_0\cdot\partial_x\vec{m}_0d^2r=\alpha\int \partial_y\vec{m}_0\cdot\partial_y\vec{m}_0d^2r$, and $F_i^{\rm SOT}=\int \vec{T}(\vec{m})\cdot(\vec{m}\times\partial_{R_i}\vec{m})d^2r$ for $\omega_D=0$. Here $F_x^{\rm SOT}$ is proportional to $\tau_{\rm d}$ and $F_y^{\rm SOT}$ is proportional to $\tau_{\rm f}$.
	
	First we calculate the skyrmion Hall angle for $\omega_D=0$, i.e., $\theta_{\rm SkHE,\pm}^{\rm N}$. After some algebra, Eq.~(\ref{Eq(S):Thiele}) leads to
	\begin{equation}
	\tan\theta_{\rm SkHE,\pm}^{\rm N}=\left.\frac{\dot{Y}}{\dot{X}}\right|_{\omega_D=0}=\frac{G F_x^{\rm SOT}+\mathcal{D}F_y^{\rm SOT}}{\mathcal{D} F_x^{\rm SOT}-GF_y^{\rm SOT}},
	\end{equation}
	which is consistent with Refs.~\cite{Jiang2016,Litzius2017}. For an opposite polarity ($\vec{m}_0\to-\vec{m}_0$), $G$ and $F_y^{\rm SOT}$ change their signs, thus $\tan\theta_{\rm SkHE,\pm}^{\rm N}$ is symmetric under the reversal of the skyrmion polarity.
	
	Now we restore a nonzero $\omega_D$. From Eq.~(\ref{Eq(S):Thiele}), we obtain after some algebra
	\begin{equation}
	\tan\theta_{\rm SkHE,\pm}^{\rm I}=\frac{\dot{Y}}{\dot{X}}=\tan\left(-\omega_D+\theta_{\rm SkHE,\pm}^{\rm N}\right),\label{Eq(S):Thiele result}
	\end{equation}
	which is consistent with Eq.~(5) in the main text.	Since Eq.~(5) is obtained without assuming any ansatz, Eq.~(\ref{Eq(S):Thiele result}) is a special case of it. Nevertheless, one can find more physical meaning from this calculation. Since we do not explicitly assume the form of the DMI here, the skyrmion Hall effect for an intermediate skyrmion is asymmetric regardless of the microscopic mechanism. For instance, if an intermediate skyrmion is stabilized by dipolar interactions (without the bulk DMI)~\cite{Buettner2018}, it also leads to an asymmetric skyrmion Hall effect.

\end{appendix}


\begin{thebibliography}{92}%
	\makeatletter
	\providecommand \@ifxundefined [1]{%
		\@ifx{#1\undefined}
	}%
	\providecommand \@ifnum [1]{%
		\ifnum #1\expandafter \@firstoftwo
		\else \expandafter \@secondoftwo
		\fi
	}%
	\providecommand \@ifx [1]{%
		\ifx #1\expandafter \@firstoftwo
		\else \expandafter \@secondoftwo
		\fi
	}%
	\providecommand \natexlab [1]{#1}%
	\providecommand \enquote  [1]{``#1''}%
	\providecommand \bibnamefont  [1]{#1}%
	\providecommand \bibfnamefont [1]{#1}%
	\providecommand \citenamefont [1]{#1}%
	\providecommand \href@noop [0]{\@secondoftwo}%
	\providecommand \href [0]{\begingroup \@sanitize@url \@href}%
	\providecommand \@href[1]{\@@startlink{#1}\@@href}%
	\providecommand \@@href[1]{\endgroup#1\@@endlink}%
	\providecommand \@sanitize@url [0]{\catcode `\\12\catcode `\$12\catcode
		`\&12\catcode `\#12\catcode `\^12\catcode `\_12\catcode `\%12\relax}%
	\providecommand \@@startlink[1]{}%
	\providecommand \@@endlink[0]{}%
	\providecommand \url  [0]{\begingroup\@sanitize@url \@url }%
	\providecommand \@url [1]{\endgroup\@href {#1}{\urlprefix }}%
	\providecommand \urlprefix  [0]{URL }%
	\providecommand \Eprint [0]{\href }%
	\providecommand \doibase [0]{http://dx.doi.org/}%
	\providecommand \selectlanguage [0]{\@gobble}%
	\providecommand \bibinfo  [0]{\@secondoftwo}%
	\providecommand \bibfield  [0]{\@secondoftwo}%
	\providecommand \translation [1]{[#1]}%
	\providecommand \BibitemOpen [0]{}%
	\providecommand \bibitemStop [0]{}%
	\providecommand \bibitemNoStop [0]{.\EOS\space}%
	\providecommand \EOS [0]{\spacefactor3000\relax}%
	\providecommand \BibitemShut  [1]{\csname bibitem#1\endcsname}%
	\let\auto@bib@innerbib\@empty
	\bibitem [{\citenamefont {Bogdanov}\ and\ \citenamefont
		{Yablonskii}(1989)}]{Bogdanov1989}%
	\BibitemOpen
	\bibfield  {author} {\bibinfo {author} {\bibfnamefont {A.}~\bibnamefont
			{Bogdanov}}\ and\ \bibinfo {author} {\bibfnamefont {D.}~\bibnamefont
			{Yablonskii}},\ }\href@noop {} {\bibfield  {journal} {\bibinfo  {journal}
			{Zh. Eksp. Teor. Fiz.}\ }\textbf {\bibinfo {volume} {95}},\ \bibinfo {pages}
		{178} (\bibinfo {year} {1989})} [JETP \textbf{68},101 (1989)]\BibitemShut {NoStop}%
	\bibitem [{\citenamefont {Bogadnov}\ \emph {et~al.}(1989)\citenamefont
		{Bogadnov}, \citenamefont {Kudinov},\ and\ \citenamefont
		{Yablonskii}}]{Bogdanov1989b}%
	\BibitemOpen
	\bibfield  {author} {\bibinfo {author} {\bibfnamefont {A.~N.}\ \bibnamefont
			{Bogadnov}}, \bibinfo {author} {\bibfnamefont {M.~V.}\ \bibnamefont
			{Kudinov}}, \ and\ \bibinfo {author} {\bibfnamefont {D.~A.}\ \bibnamefont
			{Yablonskii}},\ }\href@noop {} {\bibfield  {journal} {\bibinfo  {journal}
			{Sov. Phys. Solid State}\ }\textbf {\bibinfo {volume} {31}},\ \bibinfo
		{pages} {1707} (\bibinfo {year} {1989})} \BibitemShut {NoStop}%
	\bibitem [{\citenamefont {Bogdanov}\ and\ \citenamefont
		{Hubert}(1999)}]{Bogdanov1999}%
	\BibitemOpen
	\bibfield  {author} {\bibinfo {author} {\bibfnamefont {A.}~\bibnamefont
			{Bogdanov}}\ and\ \bibinfo {author} {\bibfnamefont {A.}~\bibnamefont
			{Hubert}},\ }\href {\doibase 10.1016/S0304-8853(98)01038-5} {\bibfield
		{journal} {\bibinfo  {journal} {J. Magn. Magn. Mater.}\ }\textbf {\bibinfo
			{volume} {195}},\ \bibinfo {pages} {182} (\bibinfo {year}
		{1999})}\BibitemShut {NoStop}%
	\bibitem [{\citenamefont {M\"{u}hlbauer}\ \emph {et~al.}(2009)\citenamefont
		{M\"{u}hlbauer}, \citenamefont {Binz}, \citenamefont {Jonietz}, \citenamefont
		{Pfleiderer}, \citenamefont {Rosch}, \citenamefont {Neubauer}, \citenamefont
		{Georgii},\ and\ \citenamefont {Boni}}]{Muhlbauer2009a}%
	\BibitemOpen
	\bibfield  {author} {\bibinfo {author} {\bibfnamefont {S.}~\bibnamefont
			{M\"{u}hlbauer}}, \bibinfo {author} {\bibfnamefont {B.}~\bibnamefont {Binz}},
		\bibinfo {author} {\bibfnamefont {F.}~\bibnamefont {Jonietz}}, \bibinfo
		{author} {\bibfnamefont {C.}~\bibnamefont {Pfleiderer}}, \bibinfo {author}
		{\bibfnamefont {A.}~\bibnamefont {Rosch}}, \bibinfo {author} {\bibfnamefont
			{A.}~\bibnamefont {Neubauer}}, \bibinfo {author} {\bibfnamefont
			{R.}~\bibnamefont {Georgii}}, \ and\ \bibinfo {author} {\bibfnamefont
			{P.}~\bibnamefont {Boni}},\ }\href {\doibase 10.1126/science.1166767}
	{\bibfield  {journal} {\bibinfo  {journal} {Science}\ }\textbf {\bibinfo
			{volume} {323}},\ \bibinfo {pages} {915} (\bibinfo {year}
		{2009})}\BibitemShut {NoStop}%
	\bibitem [{\citenamefont {Yu}\ \emph {et~al.}(2010)\citenamefont {Yu},
		\citenamefont {Onose}, \citenamefont {Kanazawa}, \citenamefont {Park},
		\citenamefont {Han}, \citenamefont {Matsui}, \citenamefont {Nagaosa},\ and\
		\citenamefont {Tokura}}]{Yu2010}%
	\BibitemOpen
	\bibfield  {author} {\bibinfo {author} {\bibfnamefont {X.~Z.}\ \bibnamefont
			{Yu}}, \bibinfo {author} {\bibfnamefont {Y.}~\bibnamefont {Onose}}, \bibinfo
		{author} {\bibfnamefont {N.}~\bibnamefont {Kanazawa}}, \bibinfo {author}
		{\bibfnamefont {J.~H.}\ \bibnamefont {Park}}, \bibinfo {author}
		{\bibfnamefont {J.~H.}\ \bibnamefont {Han}}, \bibinfo {author} {\bibfnamefont
			{Y.}~\bibnamefont {Matsui}}, \bibinfo {author} {\bibfnamefont
			{N.}~\bibnamefont {Nagaosa}}, \ and\ \bibinfo {author} {\bibfnamefont
			{Y.}~\bibnamefont {Tokura}},\ }\href {\doibase 10.1038/nature09124}
	{\bibfield  {journal} {\bibinfo  {journal} {Nature (London)}\ }\textbf {\bibinfo
			{volume} {465}},\ \bibinfo {pages} {901} (\bibinfo {year}
		{2010})}\BibitemShut {NoStop}%
	\bibitem [{\citenamefont {Romming}\ \emph {et~al.}(2013)\citenamefont
		{Romming}, \citenamefont {Hanneken}, \citenamefont {Menzel}, \citenamefont
		{Bickel}, \citenamefont {Wolter}, \citenamefont {von Bergmann}, \citenamefont
		{Kubetzka},\ and\ \citenamefont {Wiesendanger}}]{Romming2013}%
	\BibitemOpen
	\bibfield  {author} {\bibinfo {author} {\bibfnamefont {N.}~\bibnamefont
			{Romming}}, \bibinfo {author} {\bibfnamefont {C.}~\bibnamefont {Hanneken}},
		\bibinfo {author} {\bibfnamefont {M.}~\bibnamefont {Menzel}}, \bibinfo
		{author} {\bibfnamefont {J.~E.}\ \bibnamefont {Bickel}}, \bibinfo {author}
		{\bibfnamefont {B.}~\bibnamefont {Wolter}}, \bibinfo {author} {\bibfnamefont
			{K.}~\bibnamefont {von Bergmann}}, \bibinfo {author} {\bibfnamefont
			{A.}~\bibnamefont {Kubetzka}}, \ and\ \bibinfo {author} {\bibfnamefont
			{R.}~\bibnamefont {Wiesendanger}},\ }\href {\doibase 10.1126/science.1240573}
	{\bibfield  {journal} {\bibinfo  {journal} {Science}\ }\textbf {\bibinfo
			{volume} {341}},\ \bibinfo {pages} {636} (\bibinfo {year}
		{2013})}\BibitemShut {NoStop}%
	\bibitem [{\citenamefont {Everschor}\ \emph {et~al.}(2011)\citenamefont
		{Everschor}, \citenamefont {Garst}, \citenamefont {Duine},\ and\
		\citenamefont {Rosch}}]{Everschor2011}%
	\BibitemOpen
	\bibfield  {author} {\bibinfo {author} {\bibfnamefont {K.}~\bibnamefont
			{Everschor}}, \bibinfo {author} {\bibfnamefont {M.}~\bibnamefont {Garst}},
		\bibinfo {author} {\bibfnamefont {R.~A.}\ \bibnamefont {Duine}}, \ and\
		\bibinfo {author} {\bibfnamefont {A.}~\bibnamefont {Rosch}},\ }\href
	{\doibase 10.1103/PhysRevB.84.064401} {\bibfield  {journal} {\bibinfo
			{journal} {Phys. Rev. B}\ }\textbf {\bibinfo {volume} {84}},\ \bibinfo
		{pages} {064401} (\bibinfo {year} {2011})}\BibitemShut {NoStop}%
	\bibitem [{\citenamefont {Iwasaki}\ \emph
		{et~al.}(2013{\natexlab{a}})\citenamefont {Iwasaki}, \citenamefont
		{Mochizuki},\ and\ \citenamefont {Nagaosa}}]{Iwasaki2013a}%
	\BibitemOpen
	\bibfield  {author} {\bibinfo {author} {\bibfnamefont {J.}~\bibnamefont
			{Iwasaki}}, \bibinfo {author} {\bibfnamefont {M.}~\bibnamefont {Mochizuki}},
		\ and\ \bibinfo {author} {\bibfnamefont {N.}~\bibnamefont {Nagaosa}},\ }\href
	{\doibase 10.1038/nnano.2013.176} {\bibfield  {journal} {\bibinfo  {journal}
			{Nat. Nanotechnol.}\ }\textbf {\bibinfo {volume} {8}},\ \bibinfo {pages}
		{742} (\bibinfo {year} {2013}{\natexlab{a}})}\BibitemShut {NoStop}%
	\bibitem [{\citenamefont {Jiang}\ \emph {et~al.}(2016)\citenamefont {Jiang},
		\citenamefont {Zhang}, \citenamefont {Yu}, \citenamefont {Zhang},
		\citenamefont {Wang}, \citenamefont {{Benjamin Jungfleisch}}, \citenamefont
		{Pearson}, \citenamefont {Cheng}, \citenamefont {Heinonen}, \citenamefont
		{Wang}, \citenamefont {Zhou}, \citenamefont {Hoffmann},\ and\ \citenamefont
		{te~Velthuis}}]{Jiang2016}%
	\BibitemOpen
	\bibfield  {author} {\bibinfo {author} {\bibfnamefont {W.}~\bibnamefont
			{Jiang}}, \bibinfo {author} {\bibfnamefont {X.}~\bibnamefont {Zhang}},
		\bibinfo {author} {\bibfnamefont {G.}~\bibnamefont {Yu}}, \bibinfo {author}
		{\bibfnamefont {W.}~\bibnamefont {Zhang}}, \bibinfo {author} {\bibfnamefont
			{X.}~\bibnamefont {Wang}}, \bibinfo {author} {\bibfnamefont {M.}~\bibnamefont
			{{Benjamin Jungfleisch}}}, \bibinfo {author} {\bibfnamefont {J.~E.}\
			\bibnamefont {Pearson}}, \bibinfo {author} {\bibfnamefont {X.}~\bibnamefont
			{Cheng}}, \bibinfo {author} {\bibfnamefont {O.}~\bibnamefont {Heinonen}},
		\bibinfo {author} {\bibfnamefont {K.~L.}\ \bibnamefont {Wang}}, \bibinfo
		{author} {\bibfnamefont {Y.}~\bibnamefont {Zhou}}, \bibinfo {author}
		{\bibfnamefont {A.}~\bibnamefont {Hoffmann}}, \ and\ \bibinfo {author}
		{\bibfnamefont {S.~G.~E.}\ \bibnamefont {te~Velthuis}},\ }\href {\doibase
		10.1038/nphys3883} {\bibfield  {journal} {\bibinfo  {journal} {Nat. Phys.}\
		}\textbf {\bibinfo {volume} {13}},\ \bibinfo {pages} {162} (\bibinfo {year}
		{2016})}\BibitemShut {NoStop}%
	\bibitem [{\citenamefont {Litzius}\ \emph {et~al.}(2017)\citenamefont
		{Litzius}, \citenamefont {Lemesh}, \citenamefont {Kr{\"{u}}ger},
		\citenamefont {Bassirian}, \citenamefont {Caretta}, \citenamefont {Richter},
		\citenamefont {B{\"{u}}ttner}, \citenamefont {Sato}, \citenamefont
		{Tretiakov}, \citenamefont {F{\"{o}}rster}, \citenamefont {Reeve},
		\citenamefont {Weigand}, \citenamefont {Bykova}, \citenamefont {Stoll},
		\citenamefont {Sch{\"{u}}tz}, \citenamefont {Beach},\ and\ \citenamefont
		{Kl{\"{a}}ui}}]{Litzius2017}%
	\BibitemOpen
	\bibfield  {author} {\bibinfo {author} {\bibfnamefont {K.}~\bibnamefont
			{Litzius}}, \bibinfo {author} {\bibfnamefont {I.}~\bibnamefont {Lemesh}},
		\bibinfo {author} {\bibfnamefont {B.}~\bibnamefont {Kr{\"{u}}ger}}, \bibinfo
		{author} {\bibfnamefont {P.}~\bibnamefont {Bassirian}}, \bibinfo {author}
		{\bibfnamefont {L.}~\bibnamefont {Caretta}}, \bibinfo {author} {\bibfnamefont
			{K.}~\bibnamefont {Richter}}, \bibinfo {author} {\bibfnamefont
			{F.}~\bibnamefont {B{\"{u}}ttner}}, \bibinfo {author} {\bibfnamefont
			{K.}~\bibnamefont {Sato}}, \bibinfo {author} {\bibfnamefont {O.~A.}\
			\bibnamefont {Tretiakov}}, \bibinfo {author} {\bibfnamefont {J.}~\bibnamefont
			{F{\"{o}}rster}}, \bibinfo {author} {\bibfnamefont {R.~M.}\ \bibnamefont
			{Reeve}}, \bibinfo {author} {\bibfnamefont {M.}~\bibnamefont {Weigand}},
		\bibinfo {author} {\bibfnamefont {I.}~\bibnamefont {Bykova}}, \bibinfo
		{author} {\bibfnamefont {H.}~\bibnamefont {Stoll}}, \bibinfo {author}
		{\bibfnamefont {G.}~\bibnamefont {Sch{\"{u}}tz}}, \bibinfo {author}
		{\bibfnamefont {G.~S.~D.}\ \bibnamefont {Beach}}, \ and\ \bibinfo {author}
		{\bibfnamefont {M.}~\bibnamefont {Kl{\"{a}}ui}},\ }\href {\doibase
		10.1038/nphys4000} {\bibfield  {journal} {\bibinfo  {journal} {Nat. Phys.}\
		}\textbf {\bibinfo {volume} {13}},\ \bibinfo {pages} {170} (\bibinfo {year}
		{2017})}\BibitemShut {NoStop}%
	\bibitem [{\citenamefont {Bruno}\ \emph {et~al.}(2004)\citenamefont {Bruno},
		\citenamefont {Dugaev},\ and\ \citenamefont {Taillefumier}}]{Bruno2004}%
	\BibitemOpen
	\bibfield  {author} {\bibinfo {author} {\bibfnamefont {P.}~\bibnamefont
			{Bruno}}, \bibinfo {author} {\bibfnamefont {V.~K.}\ \bibnamefont {Dugaev}}, \
		and\ \bibinfo {author} {\bibfnamefont {M.}~\bibnamefont {Taillefumier}},\
	}\href {\doibase 10.1103/PhysRevLett.93.096806} {\bibfield  {journal}
		{\bibinfo  {journal} {Phys. Rev. Lett.}\ }\textbf {\bibinfo {volume} {93}},\
		\bibinfo {pages} {096806} (\bibinfo {year} {2004})}\BibitemShut {NoStop}%
	\bibitem [{\citenamefont {Neubauer}\ \emph {et~al.}(2009)\citenamefont
		{Neubauer}, \citenamefont {Pfleiderer}, \citenamefont {Binz}, \citenamefont
		{Rosch}, \citenamefont {Ritz}, \citenamefont {Niklowitz},\ and\ \citenamefont
		{B{\"{o}}ni}}]{Neubauer2009}%
	\BibitemOpen
	\bibfield  {author} {\bibinfo {author} {\bibfnamefont {A.}~\bibnamefont
			{Neubauer}}, \bibinfo {author} {\bibfnamefont {C.}~\bibnamefont
			{Pfleiderer}}, \bibinfo {author} {\bibfnamefont {B.}~\bibnamefont {Binz}},
		\bibinfo {author} {\bibfnamefont {A.}~\bibnamefont {Rosch}}, \bibinfo
		{author} {\bibfnamefont {R.}~\bibnamefont {Ritz}}, \bibinfo {author}
		{\bibfnamefont {P.~G.}\ \bibnamefont {Niklowitz}}, \ and\ \bibinfo {author}
		{\bibfnamefont {P.}~\bibnamefont {B{\"{o}}ni}},\ }\href {\doibase
		10.1103/PhysRevLett.102.186602} {\bibfield  {journal} {\bibinfo  {journal}
			{Phys. Rev. Lett.}\ }\textbf {\bibinfo {volume} {102}},\ \bibinfo {pages}
		{186602} (\bibinfo {year} {2009})}\BibitemShut {NoStop}%
	\bibitem [{\citenamefont {Schulz}\ \emph {et~al.}(2012)\citenamefont {Schulz},
		\citenamefont {Ritz}, \citenamefont {Bauer}, \citenamefont {Halder},
		\citenamefont {Wagner}, \citenamefont {Franz}, \citenamefont {Pfleiderer},
		\citenamefont {Everschor}, \citenamefont {Garst},\ and\ \citenamefont
		{Rosch}}]{Schulz2012}%
	\BibitemOpen
	\bibfield  {author} {\bibinfo {author} {\bibfnamefont {T.}~\bibnamefont
			{Schulz}}, \bibinfo {author} {\bibfnamefont {R.}~\bibnamefont {Ritz}},
		\bibinfo {author} {\bibfnamefont {A.}~\bibnamefont {Bauer}}, \bibinfo
		{author} {\bibfnamefont {M.}~\bibnamefont {Halder}}, \bibinfo {author}
		{\bibfnamefont {M.}~\bibnamefont {Wagner}}, \bibinfo {author} {\bibfnamefont
			{C.}~\bibnamefont {Franz}}, \bibinfo {author} {\bibfnamefont
			{C.}~\bibnamefont {Pfleiderer}}, \bibinfo {author} {\bibfnamefont
			{K.}~\bibnamefont {Everschor}}, \bibinfo {author} {\bibfnamefont
			{M.}~\bibnamefont {Garst}}, \ and\ \bibinfo {author} {\bibfnamefont
			{A.}~\bibnamefont {Rosch}},\ }\href {\doibase 10.1038/nphys2231} {\bibfield
		{journal} {\bibinfo  {journal} {Nat. Phys.}\ }\textbf {\bibinfo {volume}
			{8}},\ \bibinfo {pages} {301} (\bibinfo {year} {2012})}\BibitemShut {NoStop}%
	\bibitem [{\citenamefont {Everschor-Sitte}\ and\ \citenamefont
		{Sitte}(2014)}]{Everschor2014}%
	\BibitemOpen
	\bibfield  {author} {\bibinfo {author} {\bibfnamefont {K.}~\bibnamefont
			{Everschor-Sitte}}\ and\ \bibinfo {author} {\bibfnamefont {M.}~\bibnamefont
			{Sitte}},\ }\href {\doibase http://dx.doi.org/10.1063/1.4870695} {\bibfield
		{journal} {\bibinfo  {journal} {J. Appl. Phys.}\ }\textbf {\bibinfo {volume}
			{115}},\ \bibinfo {pages} {172602} (\bibinfo {year} {2014})}\BibitemShut
	{NoStop}%
	\bibitem [{\citenamefont {Dzyaloshinskii}(1957)}]{Dzyaloshinskii1957}%
	\BibitemOpen
	\bibfield  {author} {\bibinfo {author} {\bibfnamefont {I.~E.}\ \bibnamefont
			{Dzyaloshinskii}},\ }\href@noop {} {\bibfield  {journal} {\bibinfo  {journal}
			{Sov. Phys. JETP}\ }\textbf {\bibinfo {volume} {5}},\ \bibinfo {pages} {1259}
		(\bibinfo {year} {1957})}\BibitemShut {NoStop}%
	\bibitem [{\citenamefont {Moriya}(1960)}]{Moriya1960b}%
	\BibitemOpen
	\bibfield  {author} {\bibinfo {author} {\bibfnamefont {T.}~\bibnamefont
			{Moriya}},\ }\href {\doibase 10.1103/PhysRev.120.91} {\bibfield  {journal}
		{\bibinfo  {journal} {Phys. Rev.}\ }\textbf {\bibinfo {volume} {120}},\
		\bibinfo {pages} {91} (\bibinfo {year} {1960})}\BibitemShut {NoStop}%
	\bibitem [{\citenamefont {Fert}\ and\ \citenamefont {Levy}(1980)}]{Fert1980}%
	\BibitemOpen
	\bibfield  {author} {\bibinfo {author} {\bibfnamefont {A.}~\bibnamefont
			{Fert}}\ and\ \bibinfo {author} {\bibfnamefont {P.~M.}\ \bibnamefont
			{Levy}},\ }\href {\doibase 10.1103/PhysRevLett.44.1538} {\bibfield  {journal}
		{\bibinfo  {journal} {Phys. Rev. Lett.}\ }\textbf {\bibinfo {volume} {44}},\
		\bibinfo {pages} {1538} (\bibinfo {year} {1980})}\BibitemShut {NoStop}%
	\bibitem [{\citenamefont {Kim}\ \emph {et~al.}(2013)\citenamefont {Kim},
		\citenamefont {Lee}, \citenamefont {Lee},\ and\ \citenamefont
		{Stiles}}]{Kim2013b}%
	\BibitemOpen
	\bibfield  {author} {\bibinfo {author} {\bibfnamefont {K.-W.}\ \bibnamefont
			{Kim}}, \bibinfo {author} {\bibfnamefont {H.-W.}\ \bibnamefont {Lee}},
		\bibinfo {author} {\bibfnamefont {K.-J.}\ \bibnamefont {Lee}}, \ and\
		\bibinfo {author} {\bibfnamefont {M.~D.}\ \bibnamefont {Stiles}},\ }\href
	{\doibase 10.1103/PhysRevLett.111.216601} {\bibfield  {journal} {\bibinfo
			{journal} {Phys. Rev. Lett.}\ }\textbf {\bibinfo {volume} {111}},\ \bibinfo
		{pages} {216601} (\bibinfo {year} {2013})}\BibitemShut {NoStop}%
	\bibitem [{\citenamefont {Yang}\ \emph {et~al.}(2015)\citenamefont {Yang},
		\citenamefont {Thiaville}, \citenamefont {Rohart}, \citenamefont {Fert},\
		and\ \citenamefont {Chshiev}}]{Yang2015b}%
	\BibitemOpen
	\bibfield  {author} {\bibinfo {author} {\bibfnamefont {H.}~\bibnamefont
			{Yang}}, \bibinfo {author} {\bibfnamefont {A.}~\bibnamefont {Thiaville}},
		\bibinfo {author} {\bibfnamefont {S.}~\bibnamefont {Rohart}}, \bibinfo
		{author} {\bibfnamefont {A.}~\bibnamefont {Fert}}, \ and\ \bibinfo {author}
		{\bibfnamefont {M.}~\bibnamefont {Chshiev}},\ }\href {\doibase
		10.1103/PhysRevLett.115.267210} {\bibfield  {journal} {\bibinfo  {journal}
			{Phys. Rev. Lett.}\ }\textbf {\bibinfo {volume} {115}},\ \bibinfo {pages}
		{267210} (\bibinfo {year} {2015})}\BibitemShut {NoStop}%
	\bibitem [{\citenamefont {Pfleiderer}\ \emph {et~al.}(2010)\citenamefont
		{Pfleiderer}, \citenamefont {Adams}, \citenamefont {Bauer}, \citenamefont
		{Biberacher}, \citenamefont {Binz}, \citenamefont {Birkelbach}, \citenamefont
		{B{\"{o}}ni}, \citenamefont {Franz}, \citenamefont {Georgii}, \citenamefont
		{Janoschek}, \citenamefont {Jonietz}, \citenamefont {Keller}, \citenamefont
		{Ritz}, \citenamefont {M{\"{u}}hlbauer}, \citenamefont {M{\"{u}}nzer},
		\citenamefont {Neubauer}, \citenamefont {Pedersen},\ and\ \citenamefont
		{Rosch}}]{Pfleiderer2010}%
	\BibitemOpen
	\bibfield  {author} {\bibinfo {author} {\bibfnamefont {C.}~\bibnamefont
			{Pfleiderer}}, \bibinfo {author} {\bibfnamefont {T.}~\bibnamefont {Adams}},
		\bibinfo {author} {\bibfnamefont {A.}~\bibnamefont {Bauer}}, \bibinfo
		{author} {\bibfnamefont {W.}~\bibnamefont {Biberacher}}, \bibinfo {author}
		{\bibfnamefont {B.}~\bibnamefont {Binz}}, \bibinfo {author} {\bibfnamefont
			{F.}~\bibnamefont {Birkelbach}}, \bibinfo {author} {\bibfnamefont
			{P.}~\bibnamefont {B{\"{o}}ni}}, \bibinfo {author} {\bibfnamefont
			{C.}~\bibnamefont {Franz}}, \bibinfo {author} {\bibfnamefont
			{R.}~\bibnamefont {Georgii}}, \bibinfo {author} {\bibfnamefont
			{M.}~\bibnamefont {Janoschek}}, \bibinfo {author} {\bibfnamefont
			{F.}~\bibnamefont {Jonietz}}, \bibinfo {author} {\bibfnamefont
			{T.}~\bibnamefont {Keller}}, \bibinfo {author} {\bibfnamefont
			{R.}~\bibnamefont {Ritz}}, \bibinfo {author} {\bibfnamefont {S.}~\bibnamefont
			{M{\"{u}}hlbauer}}, \bibinfo {author} {\bibfnamefont {W.}~\bibnamefont
			{M{\"{u}}nzer}}, \bibinfo {author} {\bibfnamefont {A.}~\bibnamefont
			{Neubauer}}, \bibinfo {author} {\bibfnamefont {B.}~\bibnamefont {Pedersen}},
		\ and\ \bibinfo {author} {\bibfnamefont {A.}~\bibnamefont {Rosch}},\
	}\href@noop {} {\bibfield  {journal} {\bibinfo  {journal} {J. Phys.: Condens.
				Matter}\ }\textbf {\bibinfo {volume} {22}},\ \bibinfo {pages} {164207}
		(\bibinfo {year} {2010})}\BibitemShut {NoStop}%
	\bibitem [{\citenamefont {Yu}\ \emph {et~al.}(2011)\citenamefont {Yu},
		\citenamefont {Kanazawa}, \citenamefont {Onose}, \citenamefont {Kimoto},
		\citenamefont {Zhang}, \citenamefont {Ishiwata}, \citenamefont {Matsui},\
		and\ \citenamefont {Tokura}}]{Yu2011b}%
	\BibitemOpen
	\bibfield  {author} {\bibinfo {author} {\bibfnamefont {X.~Z.}\ \bibnamefont
			{Yu}}, \bibinfo {author} {\bibfnamefont {N.}~\bibnamefont {Kanazawa}},
		\bibinfo {author} {\bibfnamefont {Y.}~\bibnamefont {Onose}}, \bibinfo
		{author} {\bibfnamefont {K.}~\bibnamefont {Kimoto}}, \bibinfo {author}
		{\bibfnamefont {W.~Z.}\ \bibnamefont {Zhang}}, \bibinfo {author}
		{\bibfnamefont {S.}~\bibnamefont {Ishiwata}}, \bibinfo {author}
		{\bibfnamefont {Y.}~\bibnamefont {Matsui}}, \ and\ \bibinfo {author}
		{\bibfnamefont {Y.}~\bibnamefont {Tokura}},\ }\href@noop {} {\bibfield
		{journal} {\bibinfo  {journal} {Nat. Mater.}\ }\textbf {\bibinfo {volume}
			{10}},\ \bibinfo {pages} {106} (\bibinfo {year} {2011})}\BibitemShut
	{NoStop}%
	\bibitem [{\citenamefont {Shibata}\ \emph {et~al.}(2013)\citenamefont
		{Shibata}, \citenamefont {Yu}, \citenamefont {Hara}, \citenamefont
		{Morikawa}, \citenamefont {Kanazawa}, \citenamefont {Kimoto}, \citenamefont
		{Ishiwata}, \citenamefont {Matsui},\ and\ \citenamefont
		{Tokura}}]{Shibata2013}%
	\BibitemOpen
	\bibfield  {author} {\bibinfo {author} {\bibfnamefont {K.}~\bibnamefont
			{Shibata}}, \bibinfo {author} {\bibfnamefont {X.~Z.}\ \bibnamefont {Yu}},
		\bibinfo {author} {\bibfnamefont {T.}~\bibnamefont {Hara}}, \bibinfo {author}
		{\bibfnamefont {D.}~\bibnamefont {Morikawa}}, \bibinfo {author}
		{\bibfnamefont {N.}~\bibnamefont {Kanazawa}}, \bibinfo {author}
		{\bibfnamefont {K.}~\bibnamefont {Kimoto}}, \bibinfo {author} {\bibfnamefont
			{S.}~\bibnamefont {Ishiwata}}, \bibinfo {author} {\bibfnamefont
			{Y.}~\bibnamefont {Matsui}}, \ and\ \bibinfo {author} {\bibfnamefont
			{Y.}~\bibnamefont {Tokura}},\ }\href {\doibase 10.1038/nnano.2013.174}
	{\bibfield  {journal} {\bibinfo  {journal} {Nat. Nanotechnol.}\ }\textbf
		{\bibinfo {volume} {8}},\ \bibinfo {pages} {723} (\bibinfo {year}
		{2013})}\BibitemShut {NoStop}%
	\bibitem [{\citenamefont {Grigoriev}\ \emph {et~al.}(2013)\citenamefont
		{Grigoriev}, \citenamefont {Potapova}, \citenamefont {Siegfried},
		\citenamefont {Dyadkin}, \citenamefont {Moskvin}, \citenamefont {Dmitriev},
		\citenamefont {Menzel}, \citenamefont {Dewhurst}, \citenamefont {Chernyshov},
		\citenamefont {Sadykov}, \citenamefont {Fomicheva},\ and\ \citenamefont
		{Tsvyashchenko}}]{Grigoriev2013}%
	\BibitemOpen
	\bibfield  {author} {\bibinfo {author} {\bibfnamefont {S.~V.}\ \bibnamefont
			{Grigoriev}}, \bibinfo {author} {\bibfnamefont {N.~M.}\ \bibnamefont
			{Potapova}}, \bibinfo {author} {\bibfnamefont {S.}~\bibnamefont {Siegfried}},
		\bibinfo {author} {\bibfnamefont {V.~A.}\ \bibnamefont {Dyadkin}}, \bibinfo
		{author} {\bibfnamefont {E.~V.}\ \bibnamefont {Moskvin}}, \bibinfo {author}
		{\bibfnamefont {V.}~\bibnamefont {Dmitriev}}, \bibinfo {author}
		{\bibfnamefont {D.}~\bibnamefont {Menzel}}, \bibinfo {author} {\bibfnamefont
			{C.~D.}\ \bibnamefont {Dewhurst}}, \bibinfo {author} {\bibfnamefont
			{D.}~\bibnamefont {Chernyshov}}, \bibinfo {author} {\bibfnamefont {R.~A.}\
			\bibnamefont {Sadykov}}, \bibinfo {author} {\bibfnamefont {L.~N.}\
			\bibnamefont {Fomicheva}}, \ and\ \bibinfo {author} {\bibfnamefont {A.~V.}\
			\bibnamefont {Tsvyashchenko}},\ }\href {\doibase
		10.1103/PhysRevLett.110.207201} {\bibfield  {journal} {\bibinfo  {journal}
			{Phys. Rev. Lett.}\ }\textbf {\bibinfo {volume} {110}},\ \bibinfo {pages}
		{207201} (\bibinfo {year} {2013})}\BibitemShut {NoStop}%
	\bibitem [{\citenamefont {Gayles}\ \emph {et~al.}(2015)\citenamefont {Gayles},
		\citenamefont {Freimuth}, \citenamefont {Schena}, \citenamefont {Lani},
		\citenamefont {Mavropoulos}, \citenamefont {Duine}, \citenamefont
		{Bl{\"{u}}gel}, \citenamefont {Sinova},\ and\ \citenamefont
		{Mokrousov}}]{Gayles2015}%
	\BibitemOpen
	\bibfield  {author} {\bibinfo {author} {\bibfnamefont {J.}~\bibnamefont
			{Gayles}}, \bibinfo {author} {\bibfnamefont {F.}~\bibnamefont {Freimuth}},
		\bibinfo {author} {\bibfnamefont {T.}~\bibnamefont {Schena}}, \bibinfo
		{author} {\bibfnamefont {G.}~\bibnamefont {Lani}}, \bibinfo {author}
		{\bibfnamefont {P.}~\bibnamefont {Mavropoulos}}, \bibinfo {author}
		{\bibfnamefont {R.~A.}\ \bibnamefont {Duine}}, \bibinfo {author}
		{\bibfnamefont {S.}~\bibnamefont {Bl{\"{u}}gel}}, \bibinfo {author}
		{\bibfnamefont {J.}~\bibnamefont {Sinova}}, \ and\ \bibinfo {author}
		{\bibfnamefont {Y.}~\bibnamefont {Mokrousov}},\ }\href {\doibase
		10.1103/PhysRevLett.115.036602} {\bibfield  {journal} {\bibinfo  {journal}
			{Phys. Rev. Lett.}\ }\textbf {\bibinfo {volume} {115}},\ \bibinfo {pages}
		{036602} (\bibinfo {year} {2015})}\BibitemShut {NoStop}%
	\bibitem [{\citenamefont {{Turgut}}\ \emph {et~al.}()\citenamefont {{Turgut}},
		\citenamefont {{Paik}}, \citenamefont {{Nguyen}}, \citenamefont {{Muller}},
		\citenamefont {{Schlom}},\ and\ \citenamefont {{Fuchs}}}]{Turgut2018}%
	\BibitemOpen
	\bibfield  {author} {\bibinfo {author} {\bibfnamefont {E.}~\bibnamefont
			{{Turgut}}}, \bibinfo {author} {\bibfnamefont {H.}~\bibnamefont {{Paik}}},
		\bibinfo {author} {\bibfnamefont {K.}~\bibnamefont {{Nguyen}}}, \bibinfo
		{author} {\bibfnamefont {D.~A.}\ \bibnamefont {{Muller}}}, \bibinfo {author}
		{\bibfnamefont {D.~G.}\ \bibnamefont {{Schlom}}}, \ and\ \bibinfo {author}
		{\bibfnamefont {G.~D.}\ \bibnamefont {{Fuchs}}},\ }\href@noop {} {\ }\Eprint
	{http://arxiv.org/abs/1802.05107 (2018)} {arXiv:1802.05107}
	\BibitemShut {NoStop}%
	\bibitem [{\citenamefont {Koretsune}\ \emph {et~al.}(2015)\citenamefont
		{Koretsune}, \citenamefont {Nagaosa},\ and\ \citenamefont
		{Arita}}]{Koretsune2015}%
	\BibitemOpen
	\bibfield  {author} {\bibinfo {author} {\bibfnamefont {T.}~\bibnamefont
			{Koretsune}}, \bibinfo {author} {\bibfnamefont {N.}~\bibnamefont {Nagaosa}},
		\ and\ \bibinfo {author} {\bibfnamefont {R.}~\bibnamefont {Arita}},\
	}\href@noop {} {\bibfield  {journal} {\bibinfo  {journal} {Sci. Rep.}\
		}\textbf {\bibinfo {volume} {5}},\ \bibinfo {pages} {13302} (\bibinfo {year}
		{2015})}\BibitemShut {NoStop}%
	\bibitem [{\citenamefont {Jiang}\ \emph {et~al.}(2015)\citenamefont {Jiang},
		\citenamefont {Upadhyaya}, \citenamefont {Zhang}, \citenamefont {Yu},
		\citenamefont {Jungfleisch}, \citenamefont {Fradin}, \citenamefont {Pearson},
		\citenamefont {Tserkovnyak}, \citenamefont {Wang}, \citenamefont {Heinonen},
		\citenamefont {te~Velthuis},\ and\ \citenamefont {Hoffmann}}]{Jiang2015a}%
	\BibitemOpen
	\bibfield  {author} {\bibinfo {author} {\bibfnamefont {W.}~\bibnamefont
			{Jiang}}, \bibinfo {author} {\bibfnamefont {P.}~\bibnamefont {Upadhyaya}},
		\bibinfo {author} {\bibfnamefont {W.}~\bibnamefont {Zhang}}, \bibinfo
		{author} {\bibfnamefont {G.}~\bibnamefont {Yu}}, \bibinfo {author}
		{\bibfnamefont {M.~B.}\ \bibnamefont {Jungfleisch}}, \bibinfo {author}
		{\bibfnamefont {F.~Y.}\ \bibnamefont {Fradin}}, \bibinfo {author}
		{\bibfnamefont {J.~E.}\ \bibnamefont {Pearson}}, \bibinfo {author}
		{\bibfnamefont {Y.}~\bibnamefont {Tserkovnyak}}, \bibinfo {author}
		{\bibfnamefont {K.~L.}\ \bibnamefont {Wang}}, \bibinfo {author}
		{\bibfnamefont {O.}~\bibnamefont {Heinonen}}, \bibinfo {author}
		{\bibfnamefont {S.~G.~E.}\ \bibnamefont {te~Velthuis}}, \ and\ \bibinfo
		{author} {\bibfnamefont {A.}~\bibnamefont {Hoffmann}},\ }\href {\doibase
		10.1126/science.aaa1442} {\bibfield  {journal} {\bibinfo  {journal}
			{Science}\ }\textbf {\bibinfo {volume} {349}},\ \bibinfo {pages} {283}
		(\bibinfo {year} {2015})}\BibitemShut {NoStop}%
	\bibitem [{\citenamefont {Woo}\ \emph {et~al.}(2016)\citenamefont {Woo},
		\citenamefont {Litzius}, \citenamefont {Kr{\"{u}}ger}, \citenamefont {Im},
		\citenamefont {Caretta}, \citenamefont {Richter}, \citenamefont {Mann},
		\citenamefont {Krone}, \citenamefont {Reeve}, \citenamefont {Weigand},
		\citenamefont {Agrawal}, \citenamefont {Lemesh}, \citenamefont {Mawass},
		\citenamefont {Fischer}, \citenamefont {Kl{\"{a}}ui},\ and\ \citenamefont
		{Beach}}]{Woo2016}%
	\BibitemOpen
	\bibfield  {author} {\bibinfo {author} {\bibfnamefont {S.}~\bibnamefont
			{Woo}}, \bibinfo {author} {\bibfnamefont {K.}~\bibnamefont {Litzius}},
		\bibinfo {author} {\bibfnamefont {B.}~\bibnamefont {Kr{\"{u}}ger}}, \bibinfo
		{author} {\bibfnamefont {M.-y.}\ \bibnamefont {Im}}, \bibinfo {author}
		{\bibfnamefont {L.}~\bibnamefont {Caretta}}, \bibinfo {author} {\bibfnamefont
			{K.}~\bibnamefont {Richter}}, \bibinfo {author} {\bibfnamefont
			{M.}~\bibnamefont {Mann}}, \bibinfo {author} {\bibfnamefont {A.}~\bibnamefont
			{Krone}}, \bibinfo {author} {\bibfnamefont {R.~M.}\ \bibnamefont {Reeve}},
		\bibinfo {author} {\bibfnamefont {M.}~\bibnamefont {Weigand}}, \bibinfo
		{author} {\bibfnamefont {P.}~\bibnamefont {Agrawal}}, \bibinfo {author}
		{\bibfnamefont {I.}~\bibnamefont {Lemesh}}, \bibinfo {author} {\bibfnamefont
			{M.-A.}\ \bibnamefont {Mawass}}, \bibinfo {author} {\bibfnamefont
			{P.}~\bibnamefont {Fischer}}, \bibinfo {author} {\bibfnamefont
			{M.}~\bibnamefont {Kl{\"{a}}ui}}, \ and\ \bibinfo {author} {\bibfnamefont
			{G.~S.~D.}\ \bibnamefont {Beach}},\ }\href {\doibase 10.1038/nmat4593}
	{\bibfield  {journal} {\bibinfo  {journal} {Nat. Mater.}\ }\textbf {\bibinfo
			{volume} {15}},\ \bibinfo {pages} {501} (\bibinfo {year} {2016})}\BibitemShut
	{NoStop}%
	\bibitem [{\citenamefont {Boulle}\ \emph {et~al.}(2016)\citenamefont {Boulle},
		\citenamefont {Vogel}, \citenamefont {Yang}, \citenamefont {Pizzini},
		\citenamefont {{de Souza Chaves}}, \citenamefont {Locatelli}, \citenamefont
		{Mente\c{s}}, \citenamefont {Sala}, \citenamefont {Buda-Prejbeanu},
		\citenamefont {Klein}, \citenamefont {Belmeguenai}, \citenamefont
		{Roussign{\'{e}}}, \citenamefont {Stashkevich}, \citenamefont {Ch{\'{e}}rif},
		\citenamefont {Aballe}, \citenamefont {Foerster}, \citenamefont {Chshiev},
		\citenamefont {Auffret}, \citenamefont {Miron},\ and\ \citenamefont
		{Gaudin}}]{Boulle2016}%
	\BibitemOpen
	\bibfield  {author} {\bibinfo {author} {\bibfnamefont {O.}~\bibnamefont
			{Boulle}}, \bibinfo {author} {\bibfnamefont {J.}~\bibnamefont {Vogel}},
		\bibinfo {author} {\bibfnamefont {H.}~\bibnamefont {Yang}}, \bibinfo {author}
		{\bibfnamefont {S.}~\bibnamefont {Pizzini}}, \bibinfo {author} {\bibfnamefont
			{D.}~\bibnamefont {{de Souza Chaves}}}, \bibinfo {author} {\bibfnamefont
			{A.}~\bibnamefont {Locatelli}}, \bibinfo {author} {\bibfnamefont {T.~O.}\
			\bibnamefont {Mente\c{s}}}, \bibinfo {author} {\bibfnamefont
			{A.}~\bibnamefont {Sala}}, \bibinfo {author} {\bibfnamefont {L.~D.}\
			\bibnamefont {Buda-Prejbeanu}}, \bibinfo {author} {\bibfnamefont
			{O.}~\bibnamefont {Klein}}, \bibinfo {author} {\bibfnamefont
			{M.}~\bibnamefont {Belmeguenai}}, \bibinfo {author} {\bibfnamefont
			{Y.}~\bibnamefont {Roussign{\'{e}}}}, \bibinfo {author} {\bibfnamefont
			{A.}~\bibnamefont {Stashkevich}}, \bibinfo {author} {\bibfnamefont {S.~M.}\
			\bibnamefont {Ch{\'{e}}rif}}, \bibinfo {author} {\bibfnamefont
			{L.}~\bibnamefont {Aballe}}, \bibinfo {author} {\bibfnamefont
			{M.}~\bibnamefont {Foerster}}, \bibinfo {author} {\bibfnamefont
			{M.}~\bibnamefont {Chshiev}}, \bibinfo {author} {\bibfnamefont
			{S.}~\bibnamefont {Auffret}}, \bibinfo {author} {\bibfnamefont {I.~M.}\
			\bibnamefont {Miron}}, \ and\ \bibinfo {author} {\bibfnamefont
			{G.}~\bibnamefont {Gaudin}},\ }\href {\doibase 10.1038/nnano.2015.315}
	{\bibfield  {journal} {\bibinfo  {journal} {Nat. Nanotechnol.}\ }\textbf
		{\bibinfo {volume} {11}},\ \bibinfo {pages} {449} (\bibinfo {year}
		{2016})}\BibitemShut {NoStop}%
	\bibitem [{\citenamefont {Birss}(1964)}]{Birss1964}%
	\BibitemOpen
	\bibfield  {author} {\bibinfo {author} {\bibfnamefont {R.~R.}\ \bibnamefont
			{Birss}},\ }\href@noop {} {\emph {\bibinfo {title} {{Symmetry and
					magnetism}}}},\ Vol.\ \bibinfo {volume} {863}\ (\bibinfo  {publisher}
	{North-Holland Amsterdam},\ \bibinfo {year} {1964})\BibitemShut {NoStop}%
	\bibitem [{\citenamefont {K{\'{e}}zsm{\'{a}}rki}\ \emph
		{et~al.}(2015)\citenamefont {K{\'{e}}zsm{\'{a}}rki}, \citenamefont
		{Bord{\'{a}}cs}, \citenamefont {Milde}, \citenamefont {Neuber}, \citenamefont
		{Eng}, \citenamefont {White}, \citenamefont {R{\o}nnow}, \citenamefont
		{Dewhurst}, \citenamefont {Mochizuki}, \citenamefont {Yanai}, \citenamefont
		{Nakamura}, \citenamefont {Ehlers}, \citenamefont {Tsurkan},\ and\
		\citenamefont {Loidl}}]{Kezsmarki2015}%
	\BibitemOpen
	\bibfield  {author} {\bibinfo {author} {\bibfnamefont {I.}~\bibnamefont
			{K{\'{e}}zsm{\'{a}}rki}}, \bibinfo {author} {\bibfnamefont {S.}~\bibnamefont
			{Bord{\'{a}}cs}}, \bibinfo {author} {\bibfnamefont {P.}~\bibnamefont
			{Milde}}, \bibinfo {author} {\bibfnamefont {E.}~\bibnamefont {Neuber}},
		\bibinfo {author} {\bibfnamefont {L.~M.}\ \bibnamefont {Eng}}, \bibinfo
		{author} {\bibfnamefont {J.~S.}\ \bibnamefont {White}}, \bibinfo {author}
		{\bibfnamefont {H.~M.}\ \bibnamefont {R{\o}nnow}}, \bibinfo {author}
		{\bibfnamefont {C.~D.}\ \bibnamefont {Dewhurst}}, \bibinfo {author}
		{\bibfnamefont {M.}~\bibnamefont {Mochizuki}}, \bibinfo {author}
		{\bibfnamefont {K.}~\bibnamefont {Yanai}}, \bibinfo {author} {\bibfnamefont
			{H.}~\bibnamefont {Nakamura}}, \bibinfo {author} {\bibfnamefont
			{D.}~\bibnamefont {Ehlers}}, \bibinfo {author} {\bibfnamefont
			{V.}~\bibnamefont {Tsurkan}}, \ and\ \bibinfo {author} {\bibfnamefont
			{A.}~\bibnamefont {Loidl}},\ }\href {\doibase 10.1038/nmat4402} {\bibfield
		{journal} {\bibinfo  {journal} {Nat. Mater.}\ }\textbf {\bibinfo {volume}
			{14}},\ \bibinfo {pages} {1116} (\bibinfo {year} {2015})}\BibitemShut
	{NoStop}%
	\bibitem [{\citenamefont {Hoffmann}\ \emph {et~al.}(2017)\citenamefont
		{Hoffmann}, \citenamefont {Zimmermann}, \citenamefont {M{\"{u}}ller},
		\citenamefont {Sch{\"{u}}rhoff}, \citenamefont {Kiselev}, \citenamefont
		{Melcher},\ and\ \citenamefont {Bl{\"{u}}gel}}]{Hoffmann2017}%
	\BibitemOpen
	\bibfield  {author} {\bibinfo {author} {\bibfnamefont {M.}~\bibnamefont
			{Hoffmann}}, \bibinfo {author} {\bibfnamefont {B.}~\bibnamefont
			{Zimmermann}}, \bibinfo {author} {\bibfnamefont {G.~P.}\ \bibnamefont
			{M{\"{u}}ller}}, \bibinfo {author} {\bibfnamefont {D.}~\bibnamefont
			{Sch{\"{u}}rhoff}}, \bibinfo {author} {\bibfnamefont {N.~S.}\ \bibnamefont
			{Kiselev}}, \bibinfo {author} {\bibfnamefont {C.}~\bibnamefont {Melcher}}, \
		and\ \bibinfo {author} {\bibfnamefont {S.}~\bibnamefont {Bl{\"{u}}gel}},\
	}\href {\doibase 10.1038/s41467-017-00313-0} {\bibfield  {journal} {\bibinfo
			{journal} {Nat. Commun.}\ }\textbf {\bibinfo {volume} {8}},\ \bibinfo {pages}
		{308} (\bibinfo {year} {2017})}\BibitemShut {NoStop}%
	\bibitem [{\citenamefont {Landau}\ \emph {et~al.}(1984)\citenamefont {Landau},
		\citenamefont {Pitaevskii},\ and\ \citenamefont {Lifshitz}}]{Landau1984}%
	\BibitemOpen
	\bibfield  {author} {\bibinfo {author} {\bibfnamefont {L.~D.}\ \bibnamefont
			{Landau}}, \bibinfo {author} {\bibfnamefont {L.~P.}\ \bibnamefont
			{Pitaevskii}}, \ and\ \bibinfo {author} {\bibfnamefont {E.~M.}\ \bibnamefont
			{Lifshitz}},\ }\href@noop {} {\emph {\bibinfo {title} {{Electrodynamics of
					Continuous Media, Course of Theoretical Physics Vol. 8}}}}\ (\bibinfo
	{publisher} {Pergamon, Oxford},\ \bibinfo {year} {1984})\BibitemShut
	{NoStop}%
	\bibitem [{\citenamefont {Parkin}\ \emph {et~al.}(2008)\citenamefont {Parkin},
		\citenamefont {Hayashi},\ and\ \citenamefont {Thomas}}]{Parkin2008}%
	\BibitemOpen
	\bibfield  {author} {\bibinfo {author} {\bibfnamefont {S.~S.~P.}\
			\bibnamefont {Parkin}}, \bibinfo {author} {\bibfnamefont {M.}~\bibnamefont
			{Hayashi}}, \ and\ \bibinfo {author} {\bibfnamefont {L.}~\bibnamefont
			{Thomas}},\ }\href {\doibase 10.1126/science.1145799} {\bibfield  {journal}
		{\bibinfo  {journal} {Science}\ }\textbf {\bibinfo {volume} {320}},\ \bibinfo
		{pages} {190} (\bibinfo {year} {2008})}\BibitemShut {NoStop}%
	\bibitem [{\citenamefont {Fert}\ \emph {et~al.}(2013)\citenamefont {Fert},
		\citenamefont {Cros},\ and\ \citenamefont {Sampaio}}]{Fert2013}%
	\BibitemOpen
	\bibfield  {author} {\bibinfo {author} {\bibfnamefont {A.}~\bibnamefont
			{Fert}}, \bibinfo {author} {\bibfnamefont {V.}~\bibnamefont {Cros}}, \ and\
		\bibinfo {author} {\bibfnamefont {J.}~\bibnamefont {Sampaio}},\ }\href
	{\doibase 10.1038/nnano.2013.29} {\bibfield  {journal} {\bibinfo  {journal}
			{Nat. Nanotechnol.}\ }\textbf {\bibinfo {volume} {8}},\ \bibinfo {pages}
		{152} (\bibinfo {year} {2013})}\BibitemShut {NoStop}%
	\bibitem [{\citenamefont {Jonietz}\ \emph {et~al.}(2010)\citenamefont
		{Jonietz}, \citenamefont {Muhlbauer}, \citenamefont {Pfleiderer},
		\citenamefont {Neubauer}, \citenamefont {Munzer}, \citenamefont {Bauer},
		\citenamefont {Adams}, \citenamefont {Georgii}, \citenamefont {Boni},
		\citenamefont {Duine}, \citenamefont {Everschor}, \citenamefont {Garst},\
		and\ \citenamefont {Rosch}}]{Jonietz2010}%
	\BibitemOpen
	\bibfield  {author} {\bibinfo {author} {\bibfnamefont {F.}~\bibnamefont
			{Jonietz}}, \bibinfo {author} {\bibfnamefont {S.}~\bibnamefont {Muhlbauer}},
		\bibinfo {author} {\bibfnamefont {C.}~\bibnamefont {Pfleiderer}}, \bibinfo
		{author} {\bibfnamefont {A.}~\bibnamefont {Neubauer}}, \bibinfo {author}
		{\bibfnamefont {W.}~\bibnamefont {Munzer}}, \bibinfo {author} {\bibfnamefont
			{A.}~\bibnamefont {Bauer}}, \bibinfo {author} {\bibfnamefont
			{T.}~\bibnamefont {Adams}}, \bibinfo {author} {\bibfnamefont
			{R.}~\bibnamefont {Georgii}}, \bibinfo {author} {\bibfnamefont
			{P.}~\bibnamefont {Boni}}, \bibinfo {author} {\bibfnamefont {R.~A.}\
			\bibnamefont {Duine}}, \bibinfo {author} {\bibfnamefont {K.}~\bibnamefont
			{Everschor}}, \bibinfo {author} {\bibfnamefont {M.}~\bibnamefont {Garst}}, \
		and\ \bibinfo {author} {\bibfnamefont {A.}~\bibnamefont {Rosch}},\ }\href
	{\doibase 10.1126/science.1195709} {\bibfield  {journal} {\bibinfo  {journal}
			{Science}\ }\textbf {\bibinfo {volume} {330}},\ \bibinfo {pages} {1648}
		(\bibinfo {year} {2010})}\BibitemShut {NoStop}%
	\bibitem [{\citenamefont {Iwasaki}\ \emph
		{et~al.}(2013{\natexlab{b}})\citenamefont {Iwasaki}, \citenamefont
		{Mochizuki},\ and\ \citenamefont {Nagaosa}}]{Iwasaki2013}%
	\BibitemOpen
	\bibfield  {author} {\bibinfo {author} {\bibfnamefont {J.}~\bibnamefont
			{Iwasaki}}, \bibinfo {author} {\bibfnamefont {M.}~\bibnamefont {Mochizuki}},
		\ and\ \bibinfo {author} {\bibfnamefont {N.}~\bibnamefont {Nagaosa}},\ }\href
	{\doibase 10.1038/ncomms2442} {\bibfield  {journal} {\bibinfo  {journal}
			{Nat. Commun.}\ }\textbf {\bibinfo {volume} {4}},\ \bibinfo {pages} {1463}
		(\bibinfo {year} {2013}{\natexlab{b}})}\BibitemShut {NoStop}%
	\bibitem [{\citenamefont {{Z{\'a}zvorka}}\ \emph {et~al.}()\citenamefont
		{{Z{\'a}zvorka}}, \citenamefont {{Jakobs}}, \citenamefont {{Heinze}},
		\citenamefont {{Keil}}, \citenamefont {{Kromin}}, \citenamefont {{Jaiswal}},
		\citenamefont {{Litzius}}, \citenamefont {{Jakob}}, \citenamefont {{Virnau}},
		\citenamefont {{Pinna}}, \citenamefont {{Everschor-Sitte}}, \citenamefont
		{{Donges}}, \citenamefont {{Nowak}},\ and\ \citenamefont
		{{Kl{\"a}ui}}}]{Zazvorka2018}%
	\BibitemOpen
	\bibfield  {author} {\bibinfo {author} {\bibfnamefont {J.}~\bibnamefont
			{{Z{\'a}zvorka}}}, \bibinfo {author} {\bibfnamefont {F.}~\bibnamefont
			{{Jakobs}}}, \bibinfo {author} {\bibfnamefont {D.}~\bibnamefont {{Heinze}}},
		\bibinfo {author} {\bibfnamefont {N.}~\bibnamefont {{Keil}}}, \bibinfo
		{author} {\bibfnamefont {S.}~\bibnamefont {{Kromin}}}, \bibinfo {author}
		{\bibfnamefont {S.}~\bibnamefont {{Jaiswal}}}, \bibinfo {author}
		{\bibfnamefont {K.}~\bibnamefont {{Litzius}}}, \bibinfo {author}
		{\bibfnamefont {G.}~\bibnamefont {{Jakob}}}, \bibinfo {author} {\bibfnamefont
			{P.}~\bibnamefont {{Virnau}}}, \bibinfo {author} {\bibfnamefont
			{D.}~\bibnamefont {{Pinna}}}, \bibinfo {author} {\bibfnamefont
			{K.}~\bibnamefont {{Everschor-Sitte}}}, \bibinfo {author} {\bibfnamefont
			{A.}~\bibnamefont {{Donges}}}, \bibinfo {author} {\bibfnamefont
			{U.}~\bibnamefont {{Nowak}}}, \ and\ \bibinfo {author} {\bibfnamefont
			{M.}~\bibnamefont {{Kl{\"a}ui}}},\ }\href@noop {} {\ }\Eprint
	{http://arxiv.org/abs/arXiv:1805.05924 (2018)} {arXiv:1805.05924}
	\BibitemShut {NoStop}%
	\bibitem [{\citenamefont {Zhang}\ \emph
		{et~al.}(2016{\natexlab{a}})\citenamefont {Zhang}, \citenamefont {Zhou},\
		and\ \citenamefont {Ezawa}}]{Zhang2015b}%
	\BibitemOpen
	\bibfield  {author} {\bibinfo {author} {\bibfnamefont {X.}~\bibnamefont
			{Zhang}}, \bibinfo {author} {\bibfnamefont {Y.}~\bibnamefont {Zhou}}, \ and\
		\bibinfo {author} {\bibfnamefont {M.}~\bibnamefont {Ezawa}},\ }\href
	{\doibase 10.1038/srep24795} {\bibfield  {journal} {\bibinfo  {journal} {Sci.
				Rep.}\ }\textbf {\bibinfo {volume} {6}},\ \bibinfo {pages} {24795} (\bibinfo
		{year} {2016}{\natexlab{a}})}\BibitemShut {NoStop}%
	\bibitem [{\citenamefont {Zhang}\ \emph
		{et~al.}(2016{\natexlab{b}})\citenamefont {Zhang}, \citenamefont {Zhou},\
		and\ \citenamefont {Ezawa}}]{Zhang2016h}%
	\BibitemOpen
	\bibfield  {author} {\bibinfo {author} {\bibfnamefont {X.}~\bibnamefont
			{Zhang}}, \bibinfo {author} {\bibfnamefont {Y.}~\bibnamefont {Zhou}}, \ and\
		\bibinfo {author} {\bibfnamefont {M.}~\bibnamefont {Ezawa}},\ }\href
	{\doibase 10.1038/ncomms10293} {\bibfield  {journal} {\bibinfo  {journal}
			{Nat. Commun.}\ }\textbf {\bibinfo {volume} {7}},\ \bibinfo {pages} {10293}
		(\bibinfo {year} {2016}{\natexlab{b}})}\BibitemShut {NoStop}%
	\bibitem [{\citenamefont {Huang}\ \emph {et~al.}(2017)\citenamefont {Huang},
		\citenamefont {Zhou}, \citenamefont {Chen}, \citenamefont {Shen},
		\citenamefont {Schmid}, \citenamefont {Liu},\ and\ \citenamefont
		{Wu}}]{Huang2017}%
	\BibitemOpen
	\bibfield  {author} {\bibinfo {author} {\bibfnamefont {S.}~\bibnamefont
			{Huang}}, \bibinfo {author} {\bibfnamefont {C.}~\bibnamefont {Zhou}},
		\bibinfo {author} {\bibfnamefont {G.}~\bibnamefont {Chen}}, \bibinfo {author}
		{\bibfnamefont {H.}~\bibnamefont {Shen}}, \bibinfo {author} {\bibfnamefont
			{A.~K.}\ \bibnamefont {Schmid}}, \bibinfo {author} {\bibfnamefont
			{K.}~\bibnamefont {Liu}}, \ and\ \bibinfo {author} {\bibfnamefont
			{Y.}~\bibnamefont {Wu}},\ }\href {\doibase 10.1103/PhysRevB.96.144412}
	{\bibfield  {journal} {\bibinfo  {journal} {Phys. Rev. B}\ }\textbf {\bibinfo
			{volume} {96}},\ \bibinfo {pages} {144412} (\bibinfo {year}
		{2017})}\BibitemShut {NoStop}%
	\bibitem [{\citenamefont {Wu}\ \emph {et~al.}(2017)\citenamefont {Wu},
		\citenamefont {Miao}, \citenamefont {Sun}, \citenamefont {Wu},\ and\
		\citenamefont {Ding}}]{Wu2017a}%
	\BibitemOpen
	\bibfield  {author} {\bibinfo {author} {\bibfnamefont {H.~Z.}\ \bibnamefont
			{Wu}}, \bibinfo {author} {\bibfnamefont {B.~F.}\ \bibnamefont {Miao}},
		\bibinfo {author} {\bibfnamefont {L.}~\bibnamefont {Sun}}, \bibinfo {author}
		{\bibfnamefont {D.}~\bibnamefont {Wu}}, \ and\ \bibinfo {author}
		{\bibfnamefont {H.~F.}\ \bibnamefont {Ding}},\ }\href {\doibase
		10.1103/PhysRevB.95.174416} {\bibfield  {journal} {\bibinfo  {journal} {Phys.
				Rev. B}\ }\textbf {\bibinfo {volume} {95}},\ \bibinfo {pages} {174416}
		(\bibinfo {year} {2017})}\BibitemShut {NoStop}%
	\bibitem [{\citenamefont {Rowland}\ \emph {et~al.}(2016)\citenamefont
		{Rowland}, \citenamefont {Banerjee},\ and\ \citenamefont
		{Randeria}}]{Rowland2015}%
	\BibitemOpen
	\bibfield  {author} {\bibinfo {author} {\bibfnamefont {J.}~\bibnamefont
			{Rowland}}, \bibinfo {author} {\bibfnamefont {S.}~\bibnamefont {Banerjee}}, \
		and\ \bibinfo {author} {\bibfnamefont {M.}~\bibnamefont {Randeria}},\ }\href
	{\doibase 10.1103/PhysRevB.93.020404} {\bibfield  {journal} {\bibinfo
			{journal} {Phys. Rev. B}\ }\textbf {\bibinfo {volume} {93}},\ \bibinfo
		{pages} {020404} (\bibinfo {year} {2016})}\BibitemShut {NoStop}%
	\bibitem [{\citenamefont {Cort{\'{e}}s-Ortu{\~{n}}o}\ and\ \citenamefont
		{Landeros}(2013)}]{Cortes-Ortuno2013}%
	\BibitemOpen
	\bibfield  {author} {\bibinfo {author} {\bibfnamefont {D.}~\bibnamefont
			{Cort{\'{e}}s-Ortu{\~{n}}o}}\ and\ \bibinfo {author} {\bibfnamefont
			{P.}~\bibnamefont {Landeros}},\ }\href {\doibase
		10.1088/0953-8984/25/15/156001} {\bibfield  {journal} {\bibinfo  {journal}
			{J. Phys.: Condens. Matter}\ }\textbf {\bibinfo {volume} {25}},\ \bibinfo
		{pages} {156001} (\bibinfo {year} {2013})}\BibitemShut {NoStop}%
	\bibitem [{\citenamefont {Han}\ \emph {et~al.}(2017)\citenamefont {Han},
		\citenamefont {Richardella}, \citenamefont {Siddiqui}, \citenamefont
		{Finley}, \citenamefont {Samarth},\ and\ \citenamefont {Liu}}]{Han2017}%
	\BibitemOpen
	\bibfield  {author} {\bibinfo {author} {\bibfnamefont {J.}~\bibnamefont
			{Han}}, \bibinfo {author} {\bibfnamefont {A.}~\bibnamefont {Richardella}},
		\bibinfo {author} {\bibfnamefont {S.~A.}\ \bibnamefont {Siddiqui}}, \bibinfo
		{author} {\bibfnamefont {J.}~\bibnamefont {Finley}}, \bibinfo {author}
		{\bibfnamefont {N.}~\bibnamefont {Samarth}}, \ and\ \bibinfo {author}
		{\bibfnamefont {L.}~\bibnamefont {Liu}},\ }\href {\doibase
		10.1103/PhysRevLett.119.077702} {\bibfield  {journal} {\bibinfo  {journal}
			{Phys. Rev. Lett.}\ }\textbf {\bibinfo {volume} {119}},\ \bibinfo {pages}
		{077702} (\bibinfo {year} {2017})}\BibitemShut {NoStop}%
	\bibitem [{\citenamefont {Liu}\ \emph {et~al.}(2012)\citenamefont {Liu},
		\citenamefont {Pai}, \citenamefont {Li}, \citenamefont {Tseng}, \citenamefont
		{Ralph},\ and\ \citenamefont {Buhrman}}]{Liu2012}%
	\BibitemOpen
	\bibfield  {author} {\bibinfo {author} {\bibfnamefont {L.}~\bibnamefont
			{Liu}}, \bibinfo {author} {\bibfnamefont {C.-F.}\ \bibnamefont {Pai}},
		\bibinfo {author} {\bibfnamefont {Y.}~\bibnamefont {Li}}, \bibinfo {author}
		{\bibfnamefont {H.~W.}\ \bibnamefont {Tseng}}, \bibinfo {author}
		{\bibfnamefont {D.~C.}\ \bibnamefont {Ralph}}, \ and\ \bibinfo {author}
		{\bibfnamefont {R.~A.}\ \bibnamefont {Buhrman}},\ }\href {\doibase
		10.1126/science.1218197} {\bibfield  {journal} {\bibinfo  {journal}
			{Science}\ }\textbf {\bibinfo {volume} {336}},\ \bibinfo {pages} {555}
		(\bibinfo {year} {2012})}\BibitemShut {NoStop}%
	\bibitem [{\citenamefont {Seo}\ \emph {et~al.}(2012)\citenamefont {Seo},
		\citenamefont {Kim}, \citenamefont {Ryu}, \citenamefont {Lee},\ and\
		\citenamefont {Lee}}]{Seo2012}%
	\BibitemOpen
	\bibfield  {author} {\bibinfo {author} {\bibfnamefont {S.-M.}\ \bibnamefont
			{Seo}}, \bibinfo {author} {\bibfnamefont {K.-W.}\ \bibnamefont {Kim}},
		\bibinfo {author} {\bibfnamefont {J.}~\bibnamefont {Ryu}}, \bibinfo {author}
		{\bibfnamefont {H.-W.}\ \bibnamefont {Lee}}, \ and\ \bibinfo {author}
		{\bibfnamefont {K.-J.}\ \bibnamefont {Lee}},\ }\href {\doibase
		10.1063/1.4733674} {\bibfield  {journal} {\bibinfo  {journal} {Appl. Phys.
				Lett.}\ }\textbf {\bibinfo {volume} {101}},\ \bibinfo {pages} {022405}
		(\bibinfo {year} {2012})}\BibitemShut {NoStop}%
	\bibitem [{\citenamefont {Sinova}\ \emph {et~al.}(2015)\citenamefont {Sinova},
		\citenamefont {Valenzuela}, \citenamefont {Wunderlich}, \citenamefont
		{Back},\ and\ \citenamefont {Jungwirth}}]{Sinova2015}%
	\BibitemOpen
	\bibfield  {author} {\bibinfo {author} {\bibfnamefont {J.}~\bibnamefont
			{Sinova}}, \bibinfo {author} {\bibfnamefont {S.~O.}\ \bibnamefont
			{Valenzuela}}, \bibinfo {author} {\bibfnamefont {J.}~\bibnamefont
			{Wunderlich}}, \bibinfo {author} {\bibfnamefont {C.~H.}\ \bibnamefont
			{Back}}, \ and\ \bibinfo {author} {\bibfnamefont {T.}~\bibnamefont
			{Jungwirth}},\ }\href {\doibase 10.1103/RevModPhys.87.1213} {\bibfield
		{journal} {\bibinfo  {journal} {Rev. Mod. Phys.}\ }\textbf {\bibinfo {volume}
			{87}},\ \bibinfo {pages} {1213} (\bibinfo {year} {2015})}\BibitemShut
	{NoStop}%
	\bibitem [{\citenamefont {Mellnik}\ \emph {et~al.}(2014)\citenamefont
		{Mellnik}, \citenamefont {Lee}, \citenamefont {Richardella}, \citenamefont
		{Grab}, \citenamefont {Mintun}, \citenamefont {Fischer}, \citenamefont
		{Vaezi}, \citenamefont {Manchon}, \citenamefont {Kim}, \citenamefont
		{Samarth},\ and\ \citenamefont {Ralph}}]{Mellnik2014}%
	\BibitemOpen
	\bibfield  {author} {\bibinfo {author} {\bibfnamefont {A.~R.}\ \bibnamefont
			{Mellnik}}, \bibinfo {author} {\bibfnamefont {J.~S.}\ \bibnamefont {Lee}},
		\bibinfo {author} {\bibfnamefont {A.}~\bibnamefont {Richardella}}, \bibinfo
		{author} {\bibfnamefont {J.~L.}\ \bibnamefont {Grab}}, \bibinfo {author}
		{\bibfnamefont {P.~J.}\ \bibnamefont {Mintun}}, \bibinfo {author}
		{\bibfnamefont {M.~H.}\ \bibnamefont {Fischer}}, \bibinfo {author}
		{\bibfnamefont {A.}~\bibnamefont {Vaezi}}, \bibinfo {author} {\bibfnamefont
			{A.}~\bibnamefont {Manchon}}, \bibinfo {author} {\bibfnamefont {E.-A.}\
			\bibnamefont {Kim}}, \bibinfo {author} {\bibfnamefont {N.}~\bibnamefont
			{Samarth}}, \ and\ \bibinfo {author} {\bibfnamefont {D.~C.}\ \bibnamefont
			{Ralph}},\ }\href {\doibase 10.1038/nature13534} {\bibfield  {journal}
		{\bibinfo  {journal} {Nature (London)}\ }\textbf {\bibinfo {volume} {511}},\ \bibinfo
		{pages} {449} (\bibinfo {year} {2014})}\BibitemShut {NoStop}%
	\bibitem [{\citenamefont {Jung}\ \emph {et~al.}(2008)\citenamefont {Jung},
		\citenamefont {Kim}, \citenamefont {Lee}, \citenamefont {Lee},\ and\
		\citenamefont {Lee}}]{Jung2008}%
	\BibitemOpen
	\bibfield  {author} {\bibinfo {author} {\bibfnamefont {S.-W.}\ \bibnamefont
			{Jung}}, \bibinfo {author} {\bibfnamefont {W.}~\bibnamefont {Kim}}, \bibinfo
		{author} {\bibfnamefont {T.-D.}\ \bibnamefont {Lee}}, \bibinfo {author}
		{\bibfnamefont {K.-J.}\ \bibnamefont {Lee}}, \ and\ \bibinfo {author}
		{\bibfnamefont {H.-W.}\ \bibnamefont {Lee}},\ }\href@noop {} {\bibfield
		{journal} {\bibinfo  {journal} {Appl. Phys. Lett.}\ }\textbf {\bibinfo
			{volume} {92}},\ \bibinfo {pages} {202508} (\bibinfo {year}
		{2008})}\BibitemShut {NoStop}%
	\bibitem [{\citenamefont {Kim}\ \emph {et~al.}(2016)\citenamefont {Kim},
		\citenamefont {Lee}, \citenamefont {Lee},\ and\ \citenamefont
		{Stiles}}]{Kim2016f}%
	\BibitemOpen
	\bibfield  {author} {\bibinfo {author} {\bibfnamefont {K.-W.}\ \bibnamefont
			{Kim}}, \bibinfo {author} {\bibfnamefont {K.-J.}\ \bibnamefont {Lee}},
		\bibinfo {author} {\bibfnamefont {H.-W.}\ \bibnamefont {Lee}}, \ and\
		\bibinfo {author} {\bibfnamefont {M.~D.}\ \bibnamefont {Stiles}},\ }\href
	{\doibase 10.1103/PhysRevB.94.184402} {\bibfield  {journal} {\bibinfo
			{journal} {Phys. Rev. B}\ }\textbf {\bibinfo {volume} {94}},\ \bibinfo
		{pages} {184402} (\bibinfo {year} {2016})}\BibitemShut {NoStop}%
	\bibitem [{\citenamefont {{Chaudhary}}\ \emph {et~al.}()\citenamefont
		{{Chaudhary}}, \citenamefont {{dos Santos Dias}}, \citenamefont
		{{MacDonald}},\ and\ \citenamefont {{Lounis}}}]{Chaudhary2018}%
	\BibitemOpen
	\bibfield  {author} {\bibinfo {author} {\bibfnamefont {G.}~\bibnamefont
			{{Chaudhary}}}, \bibinfo {author} {\bibfnamefont {M.}~\bibnamefont {{dos
					Santos Dias}}}, \bibinfo {author} {\bibfnamefont {A.~H.}\ \bibnamefont
			{{MacDonald}}}, \ and\ \bibinfo {author} {\bibfnamefont {S.}~\bibnamefont
			{{Lounis}}},\ }\href@noop {} {\ }\Eprint {http://arxiv.org/abs/1801.10482
		(2018)} {arXiv:1801.10482} \BibitemShut {NoStop}%
	\bibitem [{\citenamefont {Kim}\ \emph {et~al.}(2017)\citenamefont {Kim},
		\citenamefont {Kim}, \citenamefont {Wang}, \citenamefont {Sinova},\ and\
		\citenamefont {Wu}}]{Kim2017f}%
	\BibitemOpen
	\bibfield  {author} {\bibinfo {author} {\bibfnamefont {J.}~\bibnamefont
			{Kim}}, \bibinfo {author} {\bibfnamefont {K.-W.}\ \bibnamefont {Kim}},
		\bibinfo {author} {\bibfnamefont {H.}~\bibnamefont {Wang}}, \bibinfo {author}
		{\bibfnamefont {J.}~\bibnamefont {Sinova}}, \ and\ \bibinfo {author}
		{\bibfnamefont {R.}~\bibnamefont {Wu}},\ }\href {\doibase
		10.1103/PhysRevLett.119.027201} {\bibfield  {journal} {\bibinfo  {journal}
			{Phys. Rev. Lett.}\ }\textbf {\bibinfo {volume} {119}},\ \bibinfo {pages}
		{027201} (\bibinfo {year} {2017})}\BibitemShut {NoStop}%
	\bibitem [{\citenamefont {Fanchiang}\ \emph {et~al.}(2018)\citenamefont
		{Fanchiang}, \citenamefont {Chen}, \citenamefont {Tseng}, \citenamefont
		{Chen}, \citenamefont {Cheng}, \citenamefont {Yang}, \citenamefont {Wu},
		\citenamefont {Lee}, \citenamefont {Hong},\ and\ \citenamefont
		{Kwo}}]{Fanchiang2018}%
	\BibitemOpen
	\bibfield  {author} {\bibinfo {author} {\bibfnamefont {Y.~T.}\ \bibnamefont
			{Fanchiang}}, \bibinfo {author} {\bibfnamefont {K.~H.~M.}\ \bibnamefont
			{Chen}}, \bibinfo {author} {\bibfnamefont {C.~C.}\ \bibnamefont {Tseng}},
		\bibinfo {author} {\bibfnamefont {C.~C.}\ \bibnamefont {Chen}}, \bibinfo
		{author} {\bibfnamefont {C.~K.}\ \bibnamefont {Cheng}}, \bibinfo {author}
		{\bibfnamefont {S.~R.}\ \bibnamefont {Yang}}, \bibinfo {author}
		{\bibfnamefont {C.~N.}\ \bibnamefont {Wu}}, \bibinfo {author} {\bibfnamefont
			{S.~F.}\ \bibnamefont {Lee}}, \bibinfo {author} {\bibfnamefont
			{M.}~\bibnamefont {Hong}}, \ and\ \bibinfo {author} {\bibfnamefont
			{J.}~\bibnamefont {Kwo}},\ }\href@noop {} {\bibfield  {journal} {\bibinfo
			{journal} {Nat. Commun.}\ }\textbf {\bibinfo {volume} {9}},\ \bibinfo {pages}
		{223} (\bibinfo {year} {2018})}\BibitemShut {NoStop}%
	\bibitem [{\citenamefont {Zhang}\ \emph {et~al.}(2017)\citenamefont {Zhang},
		\citenamefont {Xia}, \citenamefont {Zhou}, \citenamefont {Liu}, \citenamefont
		{Zhang},\ and\ \citenamefont {Ezawa}}]{Zhang2017h}%
	\BibitemOpen
	\bibfield  {author} {\bibinfo {author} {\bibfnamefont {X.}~\bibnamefont
			{Zhang}}, \bibinfo {author} {\bibfnamefont {J.}~\bibnamefont {Xia}}, \bibinfo
		{author} {\bibfnamefont {Y.}~\bibnamefont {Zhou}}, \bibinfo {author}
		{\bibfnamefont {X.}~\bibnamefont {Liu}}, \bibinfo {author} {\bibfnamefont
			{H.}~\bibnamefont {Zhang}}, \ and\ \bibinfo {author} {\bibfnamefont
			{M.}~\bibnamefont {Ezawa}},\ }\href@noop {} {\bibfield  {journal} {\bibinfo
			{journal} {Nat. Commun.}\ }\textbf {\bibinfo {volume} {8}},\ \bibinfo {pages}
		{1717} (\bibinfo {year} {2017})}\BibitemShut {NoStop}%
	\bibitem [{\citenamefont {Leonov}\ \emph
		{et~al.}(2016{\natexlab{a}})\citenamefont {Leonov}, \citenamefont
		{Monchesky}, \citenamefont {Romming}, \citenamefont {Kubetzka}, \citenamefont
		{Bogdanov},\ and\ \citenamefont {Wiesendanger}}]{Leonov2016a}%
	\BibitemOpen
	\bibfield  {author} {\bibinfo {author} {\bibfnamefont {A.~O.}\ \bibnamefont
			{Leonov}}, \bibinfo {author} {\bibfnamefont {T.~L.}\ \bibnamefont
			{Monchesky}}, \bibinfo {author} {\bibfnamefont {N.}~\bibnamefont {Romming}},
		\bibinfo {author} {\bibfnamefont {A.}~\bibnamefont {Kubetzka}}, \bibinfo
		{author} {\bibfnamefont {A.}~\bibnamefont {Bogdanov}}, \ and\ \bibinfo
		{author} {\bibfnamefont {R.}~\bibnamefont {Wiesendanger}},\ }\href {\doibase
		10.1088/1367-2630/18/6/065003} {\bibfield  {journal} {\bibinfo  {journal}
			{New J. Phys.}\ }\textbf {\bibinfo {volume} {18}},\ \bibinfo {pages} {065003}
		(\bibinfo {year} {2016}{\natexlab{a}})}\BibitemShut {NoStop}%
	\bibitem [{\citenamefont {B{\"{u}}ttner}\ \emph {et~al.}(2018)\citenamefont
		{B{\"{u}}ttner}, \citenamefont {Lemesh},\ and\ \citenamefont
		{Beach}}]{Buettner2018}%
	\BibitemOpen
	\bibfield  {author} {\bibinfo {author} {\bibfnamefont {F.}~\bibnamefont
			{B{\"{u}}ttner}}, \bibinfo {author} {\bibfnamefont {I.}~\bibnamefont
			{Lemesh}}, \ and\ \bibinfo {author} {\bibfnamefont {G.~S.~D.}\ \bibnamefont
			{Beach}},\ }\href@noop {} {\bibfield  {journal} {\bibinfo  {journal} {Sci.
				Rep.}\ }\textbf {\bibinfo {volume} {8}},\ \bibinfo {pages} {4464} (\bibinfo
		{year} {2018})}\BibitemShut {NoStop}%
	\bibitem [{\citenamefont {R\'ozsa}\ \emph {et~al.}(2017)\citenamefont
		{R\'ozsa}, \citenamefont {Palot\'as}, \citenamefont {De\'ak}, \citenamefont
		{Simon}, \citenamefont {Yanes}, \citenamefont {Udvardi}, \citenamefont
		{Szunyogh},\ and\ \citenamefont {Nowak}}]{Rozsa2017}%
	\BibitemOpen
	\bibfield  {author} {\bibinfo {author} {\bibfnamefont {L.}~\bibnamefont
			{R\'ozsa}}, \bibinfo {author} {\bibfnamefont {K.}~\bibnamefont {Palot\'as}},
		\bibinfo {author} {\bibfnamefont {A.}~\bibnamefont {De\'ak}}, \bibinfo
		{author} {\bibfnamefont {E.}~\bibnamefont {Simon}}, \bibinfo {author}
		{\bibfnamefont {R.}~\bibnamefont {Yanes}}, \bibinfo {author} {\bibfnamefont
			{L.}~\bibnamefont {Udvardi}}, \bibinfo {author} {\bibfnamefont
			{L.}~\bibnamefont {Szunyogh}}, \ and\ \bibinfo {author} {\bibfnamefont
			{U.}~\bibnamefont {Nowak}},\ }\href@noop {} {\bibfield  {journal} {\bibinfo
			{journal} {Phys. Rev. B}\ }\textbf {\bibinfo {volume} {95}},\ \bibinfo
		{pages} {094423} (\bibinfo {year} {2017})}\BibitemShut {NoStop}%
	\bibitem [{\citenamefont {Vida}\ \emph {et~al.}(2016)\citenamefont {Vida},
		\citenamefont {Simon}, \citenamefont {R\'ozsa}, \citenamefont {Palot\'as},\
		and\ \citenamefont {Szunyogh}}]{Vida2016}%
	\BibitemOpen
	\bibfield  {author} {\bibinfo {author} {\bibfnamefont {G.~J.}\ \bibnamefont
			{Vida}}, \bibinfo {author} {\bibfnamefont {E.}~\bibnamefont {Simon}},
		\bibinfo {author} {\bibfnamefont {L.}~\bibnamefont {R\'ozsa}}, \bibinfo
		{author} {\bibfnamefont {K.}~\bibnamefont {Palot\'as}}, \ and\ \bibinfo
		{author} {\bibfnamefont {L.}~\bibnamefont {Szunyogh}},\ }\href@noop {}
	{\bibfield  {journal} {\bibinfo  {journal} {Phys. Rev. B}\ }\textbf {\bibinfo
			{volume} {94}},\ \bibinfo {pages} {214422} (\bibinfo {year}
		{2016})}\BibitemShut {NoStop}%
	\bibitem [{\citenamefont {Tatara}\ and\ \citenamefont
		{Kohno}(2004)}]{Tatara2004}%
	\BibitemOpen
	\bibfield  {author} {\bibinfo {author} {\bibfnamefont {G.}~\bibnamefont
			{Tatara}}\ and\ \bibinfo {author} {\bibfnamefont {H.}~\bibnamefont {Kohno}},\
	}\href {\doibase 10.1103/PhysRevLett.92.086601} {\bibfield  {journal}
		{\bibinfo  {journal} {Phys. Rev. Lett.}\ }\textbf {\bibinfo {volume} {92}},\
		\bibinfo {pages} {086601} (\bibinfo {year} {2004})}\BibitemShut {NoStop}%
	\bibitem [{\citenamefont {Zhang}\ and\ \citenamefont {Li}(2004)}]{Zhang2004}%
	\BibitemOpen
	\bibfield  {author} {\bibinfo {author} {\bibfnamefont {S.}~\bibnamefont
			{Zhang}}\ and\ \bibinfo {author} {\bibfnamefont {Z.}~\bibnamefont {Li}},\
	}\href {\doibase 10.1103/PhysRevLett.93.127204} {\bibfield  {journal}
		{\bibinfo  {journal} {Phys. Rev. Lett.}\ }\textbf {\bibinfo {volume} {93}},\
		\bibinfo {pages} {127204} (\bibinfo {year} {2004})}\BibitemShut {NoStop}%
	\bibitem [{\citenamefont {Emori}\ \emph {et~al.}(2013)\citenamefont {Emori},
		\citenamefont {Bauer}, \citenamefont {Ahn}, \citenamefont {Martinez},\ and\
		\citenamefont {Beach}}]{Emori2013}%
	\BibitemOpen
	\bibfield  {author} {\bibinfo {author} {\bibfnamefont {S.}~\bibnamefont
			{Emori}}, \bibinfo {author} {\bibfnamefont {U.}~\bibnamefont {Bauer}},
		\bibinfo {author} {\bibfnamefont {S.-M.}\ \bibnamefont {Ahn}}, \bibinfo
		{author} {\bibfnamefont {E.}~\bibnamefont {Martinez}}, \ and\ \bibinfo
		{author} {\bibfnamefont {G.~S.~D.}\ \bibnamefont {Beach}},\ }\href {\doibase
		10.1038/nmat3675} {\bibfield  {journal} {\bibinfo  {journal} {Nat. Mater.}\
		}\textbf {\bibinfo {volume} {12}},\ \bibinfo {pages} {611} (\bibinfo {year}
		{2013})}\BibitemShut {NoStop}%
	\bibitem [{\citenamefont {Ryu}\ \emph {et~al.}(2013)\citenamefont {Ryu},
		\citenamefont {Thomas}, \citenamefont {Yang},\ and\ \citenamefont
		{Parkin}}]{Ryu2013}%
	\BibitemOpen
	\bibfield  {author} {\bibinfo {author} {\bibfnamefont {K.-S.}\ \bibnamefont
			{Ryu}}, \bibinfo {author} {\bibfnamefont {L.}~\bibnamefont {Thomas}},
		\bibinfo {author} {\bibfnamefont {S.-H.}\ \bibnamefont {Yang}}, \ and\
		\bibinfo {author} {\bibfnamefont {S.}~\bibnamefont {Parkin}},\ }\href
	{\doibase 10.1038/nnano.2013.102} {\bibfield  {journal} {\bibinfo  {journal}
			{Nat. Nanotechnol.}\ }\textbf {\bibinfo {volume} {8}},\ \bibinfo {pages}
		{527} (\bibinfo {year} {2013})}\BibitemShut {NoStop}%
	\bibitem [{\citenamefont {Manchon}\ and\ \citenamefont
		{Zhang}(2008)}]{Manchon2008}%
	\BibitemOpen
	\bibfield  {author} {\bibinfo {author} {\bibfnamefont {A.}~\bibnamefont
			{Manchon}}\ and\ \bibinfo {author} {\bibfnamefont {S.}~\bibnamefont
			{Zhang}},\ }\href@noop {} {\bibfield  {journal} {\bibinfo  {journal} {Phys.
				Rev. B}\ }\textbf {\bibinfo {volume} {78}},\ \bibinfo {pages} {212405}
		(\bibinfo {year} {2008})}\BibitemShut {NoStop}%
	\bibitem [{\citenamefont {Matos-Abiague}\ and\ \citenamefont
		{Rodriguez-Su{\'{a}}rez}(2009)}]{Matos-Abiague2009c}%
	\BibitemOpen
	\bibfield  {author} {\bibinfo {author} {\bibfnamefont {A.}~\bibnamefont
			{Matos-Abiague}}\ and\ \bibinfo {author} {\bibfnamefont {R.~L.}\ \bibnamefont
			{Rodriguez-Su{\'{a}}rez}},\ }\href {\doibase 10.1103/PhysRevB.80.094424}
	{\bibfield  {journal} {\bibinfo  {journal} {Phys. Rev. B}\ }\textbf {\bibinfo
			{volume} {80}},\ \bibinfo {pages} {094424} (\bibinfo {year}
		{2009})}\BibitemShut {NoStop}%
	\bibitem [{\citenamefont {Wang}\ and\ \citenamefont
		{Manchon}(2012)}]{Wang2012b}%
	\BibitemOpen
	\bibfield  {author} {\bibinfo {author} {\bibfnamefont {X.}~\bibnamefont
			{Wang}}\ and\ \bibinfo {author} {\bibfnamefont {A.}~\bibnamefont {Manchon}},\
	}\href {\doibase 10.1103/PhysRevLett.108.117201} {\bibfield  {journal}
		{\bibinfo  {journal} {Phys. Rev. Lett.}\ }\textbf {\bibinfo {volume} {108}},\
		\bibinfo {pages} {117201} (\bibinfo {year} {2012})}\BibitemShut {NoStop}%
	\bibitem [{\citenamefont {Kim}\ \emph {et~al.}(2012)\citenamefont {Kim},
		\citenamefont {Seo}, \citenamefont {Ryu}, \citenamefont {Lee},\ and\
		\citenamefont {Lee}}]{Kim2012b}%
	\BibitemOpen
	\bibfield  {author} {\bibinfo {author} {\bibfnamefont {K.-W.}\ \bibnamefont
			{Kim}}, \bibinfo {author} {\bibfnamefont {S.-M.}\ \bibnamefont {Seo}},
		\bibinfo {author} {\bibfnamefont {J.}~\bibnamefont {Ryu}}, \bibinfo {author}
		{\bibfnamefont {K.-J.}\ \bibnamefont {Lee}}, \ and\ \bibinfo {author}
		{\bibfnamefont {H.-W.}\ \bibnamefont {Lee}},\ }\href {\doibase
		10.1103/PhysRevB.85.180404} {\bibfield  {journal} {\bibinfo  {journal} {Phys.
				Rev. B}\ }\textbf {\bibinfo {volume} {85}},\ \bibinfo {pages} {180404}
		(\bibinfo {year} {2012})}\BibitemShut {NoStop}%
	\bibitem [{\citenamefont {Pesin}\ and\ \citenamefont
		{MacDonald}(2012)}]{Pesin2012}%
	\BibitemOpen
	\bibfield  {author} {\bibinfo {author} {\bibfnamefont {D.~A.}\ \bibnamefont
			{Pesin}}\ and\ \bibinfo {author} {\bibfnamefont {A.~H.}\ \bibnamefont
			{MacDonald}},\ }\href {\doibase 10.1103/PhysRevB.86.014416} {\bibfield
		{journal} {\bibinfo  {journal} {Phys. Rev. B}\ }\textbf {\bibinfo {volume}
			{86}},\ \bibinfo {pages} {014416} (\bibinfo {year} {2012})}\BibitemShut
	{NoStop}%
	\bibitem [{\citenamefont {Kurebayashi}\ \emph {et~al.}(2014)\citenamefont
		{Kurebayashi}, \citenamefont {Sinova}, \citenamefont {Fang}, \citenamefont
		{Irvine}, \citenamefont {Skinner}, \citenamefont {Wunderlich}, \citenamefont
		{Nov{\'{a}}k}, \citenamefont {Campion}, \citenamefont {Gallagher},
		\citenamefont {Vehstedt}, \citenamefont {Z{\^{a}}rbo}, \citenamefont
		{V{\'{y}}born{\'{y}}}, \citenamefont {Ferguson},\ and\ \citenamefont
		{Jungwirth}}]{Kurebayashi2014}%
	\BibitemOpen
	\bibfield  {author} {\bibinfo {author} {\bibfnamefont {H.}~\bibnamefont
			{Kurebayashi}}, \bibinfo {author} {\bibfnamefont {J.}~\bibnamefont {Sinova}},
		\bibinfo {author} {\bibfnamefont {D.}~\bibnamefont {Fang}}, \bibinfo {author}
		{\bibfnamefont {A.~C.}\ \bibnamefont {Irvine}}, \bibinfo {author}
		{\bibfnamefont {T.~D.}\ \bibnamefont {Skinner}}, \bibinfo {author}
		{\bibfnamefont {J.}~\bibnamefont {Wunderlich}}, \bibinfo {author}
		{\bibfnamefont {V.}~\bibnamefont {Nov{\'{a}}k}}, \bibinfo {author}
		{\bibfnamefont {R.~P.}\ \bibnamefont {Campion}}, \bibinfo {author}
		{\bibfnamefont {B.~L.}\ \bibnamefont {Gallagher}}, \bibinfo {author}
		{\bibfnamefont {E.~K.}\ \bibnamefont {Vehstedt}}, \bibinfo {author}
		{\bibfnamefont {L.~P.}\ \bibnamefont {Z{\^{a}}rbo}}, \bibinfo {author}
		{\bibfnamefont {K.}~\bibnamefont {V{\'{y}}born{\'{y}}}}, \bibinfo {author}
		{\bibfnamefont {A.~J.}\ \bibnamefont {Ferguson}}, \ and\ \bibinfo {author}
		{\bibfnamefont {T.}~\bibnamefont {Jungwirth}},\ }\href {\doibase
		10.1038/nnano.2014.15} {\bibfield  {journal} {\bibinfo  {journal} {Nat.
				Nanotechnol.}\ }\textbf {\bibinfo {volume} {9}},\ \bibinfo {pages} {211}
		(\bibinfo {year} {2014})}\BibitemShut {NoStop}%
	\bibitem [{\citenamefont {Lee}\ \emph {et~al.}(2018)\citenamefont {Lee},
		\citenamefont {Kim}, \citenamefont {Lee},\ and\ \citenamefont
		{Lee}}]{Lee2017}%
	\BibitemOpen
	\bibfield  {author} {\bibinfo {author} {\bibfnamefont {S.-J.}\ \bibnamefont
			{Lee}}, \bibinfo {author} {\bibfnamefont {K.-W.}\ \bibnamefont {Kim}},
		\bibinfo {author} {\bibfnamefont {H.-W.}\ \bibnamefont {Lee}}, \ and\
		\bibinfo {author} {\bibfnamefont {K.-J.}\ \bibnamefont {Lee}},\ }\href@noop
	{} {\bibfield  {journal} {\bibinfo  {journal} {J. Mag. Mag. Mater.}\ }\textbf
		{\bibinfo {volume} {455}},\ \bibinfo {pages} {14} (\bibinfo {year}
		{2018})}\BibitemShut {NoStop}%
	\bibitem [{\citenamefont {Thiele}(1973)}]{Thiele1973}%
	\BibitemOpen
	\bibfield  {author} {\bibinfo {author} {\bibfnamefont {A.~A.}\ \bibnamefont
			{Thiele}},\ }\href@noop {} {\bibfield  {journal} {\bibinfo  {journal} {Phys.
				Rev. Lett.}\ }\textbf {\bibinfo {volume} {30}},\ \bibinfo {pages} {230}
		(\bibinfo {year} {1973})}\BibitemShut {NoStop}%
	\bibitem [{\citenamefont {Dhital}\ \emph {et~al.}(2017)\citenamefont {Dhital},
		\citenamefont {DeBeer-Schmitt}, \citenamefont {Zhang}, \citenamefont {Xie},
		\citenamefont {Young},\ and\ \citenamefont {DiTusa}}]{Dhital2017}%
	\BibitemOpen
	\bibfield  {author} {\bibinfo {author} {\bibfnamefont {C.}~\bibnamefont
			{Dhital}}, \bibinfo {author} {\bibfnamefont {L.}~\bibnamefont
			{DeBeer-Schmitt}}, \bibinfo {author} {\bibfnamefont {Q.}~\bibnamefont
			{Zhang}}, \bibinfo {author} {\bibfnamefont {W.}~\bibnamefont {Xie}}, \bibinfo
		{author} {\bibfnamefont {D.~P.}\ \bibnamefont {Young}}, \ and\ \bibinfo
		{author} {\bibfnamefont {J.~F.}\ \bibnamefont {DiTusa}},\ }\href@noop {}
	{\bibfield  {journal} {\bibinfo  {journal} {Phys. Rev. B}\ }\textbf {\bibinfo
			{volume} {96}},\ \bibinfo {pages} {214425} (\bibinfo {year}
		{2017})}\BibitemShut {NoStop}%
	\bibitem [{\citenamefont {Wilhelm}\ \emph {et~al.}(2012)\citenamefont
		{Wilhelm}, \citenamefont {Baenitz}, \citenamefont {Schmidt}, \citenamefont
		{Naylor}, \citenamefont {Lortz}, \citenamefont {R{\"{o}}{\ss}ler},
		\citenamefont {Leonov},\ and\ \citenamefont {Bogdanov}}]{Wilhelm}%
	\BibitemOpen
	\bibfield  {author} {\bibinfo {author} {\bibfnamefont {H.}~\bibnamefont
			{Wilhelm}}, \bibinfo {author} {\bibfnamefont {M.}~\bibnamefont {Baenitz}},
		\bibinfo {author} {\bibfnamefont {M.}~\bibnamefont {Schmidt}}, \bibinfo
		{author} {\bibfnamefont {C.}~\bibnamefont {Naylor}}, \bibinfo {author}
		{\bibfnamefont {R.}~\bibnamefont {Lortz}}, \bibinfo {author} {\bibfnamefont
			{U.~K.}\ \bibnamefont {R{\"{o}}{\ss}ler}}, \bibinfo {author} {\bibfnamefont
			{a.~a.}\ \bibnamefont {Leonov}}, \ and\ \bibinfo {author} {\bibfnamefont
			{A.}~\bibnamefont {Bogdanov}},\ }\href {\doibase
		10.1088/0953-8984/24/29/294204} {\bibfield  {journal} {\bibinfo  {journal}
			{J. Phys.: Condens. Matter}\ }\textbf {\bibinfo {volume} {24}},\ \bibinfo
		{pages} {294204} (\bibinfo {year} {2012})}\BibitemShut {NoStop}%
	\bibitem [{\citenamefont {Emori}\ \emph {et~al.}(2014)\citenamefont {Emori},
		\citenamefont {Martinez}, \citenamefont {Lee}, \citenamefont {Lee},
		\citenamefont {Bauer}, \citenamefont {Ahn}, \citenamefont {Agrawal},
		\citenamefont {Bono},\ and\ \citenamefont {Beach}}]{Emori2014}%
	\BibitemOpen
	\bibfield  {author} {\bibinfo {author} {\bibfnamefont {S.}~\bibnamefont
			{Emori}}, \bibinfo {author} {\bibfnamefont {E.}~\bibnamefont {Martinez}},
		\bibinfo {author} {\bibfnamefont {K.-J.}\ \bibnamefont {Lee}}, \bibinfo
		{author} {\bibfnamefont {H.-W.}\ \bibnamefont {Lee}}, \bibinfo {author}
		{\bibfnamefont {U.}~\bibnamefont {Bauer}}, \bibinfo {author} {\bibfnamefont
			{S.-M.}\ \bibnamefont {Ahn}}, \bibinfo {author} {\bibfnamefont
			{P.}~\bibnamefont {Agrawal}}, \bibinfo {author} {\bibfnamefont {D.~C.}\
			\bibnamefont {Bono}}, \ and\ \bibinfo {author} {\bibfnamefont {G.~S.~D.}\
			\bibnamefont {Beach}},\ }\href {\doibase 10.1103/PhysRevB.90.184427}
	{\bibfield  {journal} {\bibinfo  {journal} {Phys. Rev. B}\ }\textbf {\bibinfo
			{volume} {90}},\ \bibinfo {pages} {184427} (\bibinfo {year}
		{2014})}\BibitemShut {NoStop}%
	\bibitem [{\citenamefont {Heinonen}\ \emph {et~al.}(2016)\citenamefont
		{Heinonen}, \citenamefont {Jiang}, \citenamefont {Somaily}, \citenamefont
		{te~Velthuis},\ and\ \citenamefont {Hoffmann}}]{Heinonen2016}%
	\BibitemOpen
	\bibfield  {author} {\bibinfo {author} {\bibfnamefont {O.}~\bibnamefont
			{Heinonen}}, \bibinfo {author} {\bibfnamefont {W.}~\bibnamefont {Jiang}},
		\bibinfo {author} {\bibfnamefont {H.}~\bibnamefont {Somaily}}, \bibinfo
		{author} {\bibfnamefont {S.~G.~E.}\ \bibnamefont {te~Velthuis}}, \ and\
		\bibinfo {author} {\bibfnamefont {A.}~\bibnamefont {Hoffmann}},\ }\href
	{\doibase 10.1103/PhysRevB.93.094407} {\bibfield  {journal} {\bibinfo
			{journal} {Phys. Rev. B}\ }\textbf {\bibinfo {volume} {93}},\ \bibinfo
		{pages} {094407} (\bibinfo {year} {2016})}\BibitemShut {NoStop}%
	\bibitem [{\citenamefont {Moreau-Luchaire}\ \emph {et~al.}(2016)\citenamefont
		{Moreau-Luchaire}, \citenamefont {Moutafis}, \citenamefont {Reyren},
		\citenamefont {Sampaio}, \citenamefont {Vaz}, \citenamefont {{Van Horne}},
		\citenamefont {Bouzehouane}, \citenamefont {Garcia}, \citenamefont
		{Deranlot}, \citenamefont {Warnicke}, \citenamefont {Wohlh{\"{u}}ter},
		\citenamefont {George}, \citenamefont {Weigand}, \citenamefont {Raabe},
		\citenamefont {Cros},\ and\ \citenamefont {Fert}}]{Moreau-Luchaire2016}%
	\BibitemOpen
	\bibfield  {author} {\bibinfo {author} {\bibfnamefont {C.}~\bibnamefont
			{Moreau-Luchaire}}, \bibinfo {author} {\bibfnamefont {C.}~\bibnamefont
			{Moutafis}}, \bibinfo {author} {\bibfnamefont {N.}~\bibnamefont {Reyren}},
		\bibinfo {author} {\bibfnamefont {J.}~\bibnamefont {Sampaio}}, \bibinfo
		{author} {\bibfnamefont {C.~a.~F.}\ \bibnamefont {Vaz}}, \bibinfo {author}
		{\bibfnamefont {N.}~\bibnamefont {{Van Horne}}}, \bibinfo {author}
		{\bibfnamefont {K.}~\bibnamefont {Bouzehouane}}, \bibinfo {author}
		{\bibfnamefont {K.}~\bibnamefont {Garcia}}, \bibinfo {author} {\bibfnamefont
			{C.}~\bibnamefont {Deranlot}}, \bibinfo {author} {\bibfnamefont
			{P.}~\bibnamefont {Warnicke}}, \bibinfo {author} {\bibfnamefont
			{P.}~\bibnamefont {Wohlh{\"{u}}ter}}, \bibinfo {author} {\bibfnamefont
			{J.-M.}\ \bibnamefont {George}}, \bibinfo {author} {\bibfnamefont
			{M.}~\bibnamefont {Weigand}}, \bibinfo {author} {\bibfnamefont
			{J.}~\bibnamefont {Raabe}}, \bibinfo {author} {\bibfnamefont
			{V.}~\bibnamefont {Cros}}, \ and\ \bibinfo {author} {\bibfnamefont
			{A.}~\bibnamefont {Fert}},\ }\href {\doibase 10.1038/nnano.2015.313}
	{\bibfield  {journal} {\bibinfo  {journal} {Nat. Nanotechnol.}\ }\textbf
		{\bibinfo {volume} {11}},\ \bibinfo {pages} {444} (\bibinfo {year}
		{2016})}\BibitemShut {NoStop}%
	\bibitem [{\citenamefont {Pizzini}\ \emph {et~al.}(2014)\citenamefont
		{Pizzini}, \citenamefont {Vogel}, \citenamefont {Rohart}, \citenamefont
		{Buda-Prejbeanu}, \citenamefont {Ju{\'{e}}}, \citenamefont {Boulle},
		\citenamefont {Miron}, \citenamefont {Safeer}, \citenamefont {Auffret},
		\citenamefont {Gaudin},\ and\ \citenamefont {Thiaville}}]{Pizzini2014}%
	\BibitemOpen
	\bibfield  {author} {\bibinfo {author} {\bibfnamefont {S.}~\bibnamefont
			{Pizzini}}, \bibinfo {author} {\bibfnamefont {J.}~\bibnamefont {Vogel}},
		\bibinfo {author} {\bibfnamefont {S.}~\bibnamefont {Rohart}}, \bibinfo
		{author} {\bibfnamefont {L.~D.}\ \bibnamefont {Buda-Prejbeanu}}, \bibinfo
		{author} {\bibfnamefont {{\'{E}}.}~\bibnamefont {Ju{\'{e}}}}, \bibinfo
		{author} {\bibfnamefont {O.}~\bibnamefont {Boulle}}, \bibinfo {author}
		{\bibfnamefont {I.~M.}\ \bibnamefont {Miron}}, \bibinfo {author}
		{\bibfnamefont {C.~K.}\ \bibnamefont {Safeer}}, \bibinfo {author}
		{\bibfnamefont {S.}~\bibnamefont {Auffret}}, \bibinfo {author} {\bibfnamefont
			{G.}~\bibnamefont {Gaudin}}, \ and\ \bibinfo {author} {\bibfnamefont
			{A.}~\bibnamefont {Thiaville}},\ }\href {\doibase
		10.1103/PhysRevLett.113.047203} {\bibfield  {journal} {\bibinfo  {journal}
			{Phys. Rev. Lett.}\ }\textbf {\bibinfo {volume} {113}},\ \bibinfo {pages}
		{047203} (\bibinfo {year} {2014})}\BibitemShut {NoStop}%
	\bibitem [{\citenamefont {Belmeguenai}\ \emph {et~al.}(2015)\citenamefont
		{Belmeguenai}, \citenamefont {Adam}, \citenamefont {Roussign{\'{e}}},
		\citenamefont {Eimer}, \citenamefont {Devolder}, \citenamefont {Kim},
		\citenamefont {Cherif}, \citenamefont {Stashkevich},\ and\ \citenamefont
		{Thiaville}}]{Belmeguenai2015}%
	\BibitemOpen
	\bibfield  {author} {\bibinfo {author} {\bibfnamefont {M.}~\bibnamefont
			{Belmeguenai}}, \bibinfo {author} {\bibfnamefont {J.-p.}\ \bibnamefont
			{Adam}}, \bibinfo {author} {\bibfnamefont {Y.}~\bibnamefont
			{Roussign{\'{e}}}}, \bibinfo {author} {\bibfnamefont {S.}~\bibnamefont
			{Eimer}}, \bibinfo {author} {\bibfnamefont {T.}~\bibnamefont {Devolder}},
		\bibinfo {author} {\bibfnamefont {J.~V.}\ \bibnamefont {Kim}}, \bibinfo
		{author} {\bibfnamefont {S.~M.}\ \bibnamefont {Cherif}}, \bibinfo {author}
		{\bibfnamefont {A.}~\bibnamefont {Stashkevich}}, \ and\ \bibinfo {author}
		{\bibfnamefont {A.}~\bibnamefont {Thiaville}},\ }\href {\doibase
		10.1103/PhysRevB.91.180405} {\bibfield  {journal} {\bibinfo  {journal} {Phys.
				Rev. B}\ }\textbf {\bibinfo {volume} {91}},\ \bibinfo {pages} {180405}
		(\bibinfo {year} {2015})}\BibitemShut {NoStop}%
	\bibitem [{\citenamefont {Vansteenkiste}\ \emph {et~al.}(2014)\citenamefont
		{Vansteenkiste}, \citenamefont {Leliaert}, \citenamefont {Dvornik},
		\citenamefont {Helsen}, \citenamefont {Garcia-Sanchez},\ and\ \citenamefont
		{{Van Waeyenberge}}}]{Vansteenkiste2014}%
	\BibitemOpen
	\bibfield  {author} {\bibinfo {author} {\bibfnamefont {A.}~\bibnamefont
			{Vansteenkiste}}, \bibinfo {author} {\bibfnamefont {J.}~\bibnamefont
			{Leliaert}}, \bibinfo {author} {\bibfnamefont {M.}~\bibnamefont {Dvornik}},
		\bibinfo {author} {\bibfnamefont {M.}~\bibnamefont {Helsen}}, \bibinfo
		{author} {\bibfnamefont {F.}~\bibnamefont {Garcia-Sanchez}}, \ and\ \bibinfo
		{author} {\bibfnamefont {B.}~\bibnamefont {{Van Waeyenberge}}},\ }\href
	{\doibase 10.1063/1.4899186} {\bibfield  {journal} {\bibinfo  {journal} {AIP
				Adv.}\ }\textbf {\bibinfo {volume} {4}},\ \bibinfo {pages} {107133} (\bibinfo
		{year} {2014})}\BibitemShut {NoStop}%
	\bibitem [{\citenamefont {Reichhardt}\ \emph {et~al.}(2015)\citenamefont
		{Reichhardt}, \citenamefont {Ray},\ and\ \citenamefont
		{Reichhardt}}]{Reichhardt2015}%
	\BibitemOpen
	\bibfield  {author} {\bibinfo {author} {\bibfnamefont {C.}~\bibnamefont
			{Reichhardt}}, \bibinfo {author} {\bibfnamefont {D.}~\bibnamefont {Ray}}, \
		and\ \bibinfo {author} {\bibfnamefont {C.~J.~O.}\ \bibnamefont
			{Reichhardt}},\ }\href {\doibase 10.1103/PhysRevLett.114.217202} {\bibfield
		{journal} {\bibinfo  {journal} {Phys. Rev. Lett.}\ }\textbf {\bibinfo
			{volume} {114}},\ \bibinfo {pages} {217202} (\bibinfo {year}
		{2015})}\BibitemShut {NoStop}%
	\bibitem [{\citenamefont {{Cao}}\ \emph {et~al.}()\citenamefont {{Cao}},
		\citenamefont {{Zhang}}, \citenamefont {{Koopmans}}, \citenamefont {{Peng}},
		\citenamefont {{Zhang}}, \citenamefont {{Wang}}, \citenamefont {{Yan}},
		\citenamefont {{Yang}},\ and\ \citenamefont {{Zhao}}}]{Cao2017}%
	\BibitemOpen
	\bibfield  {author} {\bibinfo {author} {\bibfnamefont {A.}~\bibnamefont
			{{Cao}}}, \bibinfo {author} {\bibfnamefont {X.}~\bibnamefont {{Zhang}}},
		\bibinfo {author} {\bibfnamefont {B.}~\bibnamefont {{Koopmans}}}, \bibinfo
		{author} {\bibfnamefont {S.}~\bibnamefont {{Peng}}}, \bibinfo {author}
		{\bibfnamefont {Y.}~\bibnamefont {{Zhang}}}, \bibinfo {author} {\bibfnamefont
			{Z.}~\bibnamefont {{Wang}}}, \bibinfo {author} {\bibfnamefont
			{S.}~\bibnamefont {{Yan}}}, \bibinfo {author} {\bibfnamefont
			{H.}~\bibnamefont {{Yang}}}, \ and\ \bibinfo {author} {\bibfnamefont
			{W.}~\bibnamefont {{Zhao}}},\ }\href@noop {} {\ }\Eprint
	{http://arxiv.org/abs/1710.09051 (2017)} {Nanoscale (2018), doi:10.1039/C7NR08085A}
	\BibitemShut {NoStop}%
	\bibitem [{\citenamefont {Belabbes}\ \emph {et~al.}(2016)\citenamefont
		{Belabbes}, \citenamefont {Bihlmayer}, \citenamefont {Bl{\"{u}}gel},\ and\
		\citenamefont {Manchon}}]{Belabbes2015}%
	\BibitemOpen
	\bibfield  {author} {\bibinfo {author} {\bibfnamefont {A.}~\bibnamefont
			{Belabbes}}, \bibinfo {author} {\bibfnamefont {G.}~\bibnamefont {Bihlmayer}},
		\bibinfo {author} {\bibfnamefont {S.}~\bibnamefont {Bl{\"{u}}gel}}, \ and\
		\bibinfo {author} {\bibfnamefont {A.}~\bibnamefont {Manchon}},\ }\href
	{\doibase 10.1038/srep24634} {\bibfield  {journal} {\bibinfo  {journal} {Sci.
				Rep.}\ }\textbf {\bibinfo {volume} {6}},\ \bibinfo {pages} {24634} (\bibinfo
		{year} {2016})}\BibitemShut {NoStop}%
	\bibitem [{\citenamefont {Balk}\ \emph {et~al.}(2017)\citenamefont {Balk},
		\citenamefont {Kim}, \citenamefont {Pierce}, \citenamefont {Stiles},
		\citenamefont {Unguris},\ and\ \citenamefont {Stavis}}]{Balk2017}%
	\BibitemOpen
	\bibfield  {author} {\bibinfo {author} {\bibfnamefont {A.~L.}\ \bibnamefont
			{Balk}}, \bibinfo {author} {\bibfnamefont {K.-W.}\ \bibnamefont {Kim}},
		\bibinfo {author} {\bibfnamefont {D.~T.}\ \bibnamefont {Pierce}}, \bibinfo
		{author} {\bibfnamefont {M.~D.}\ \bibnamefont {Stiles}}, \bibinfo {author}
		{\bibfnamefont {J.}~\bibnamefont {Unguris}}, \ and\ \bibinfo {author}
		{\bibfnamefont {S.~M.}\ \bibnamefont {Stavis}},\ }\href {\doibase
		10.1103/PhysRevLett.119.077205} {\bibfield  {journal} {\bibinfo  {journal}
			{Phys. Rev. Lett.}\ }\textbf {\bibinfo {volume} {119}},\ \bibinfo {pages}
		{077205} (\bibinfo {year} {2017})}\BibitemShut {NoStop}%
	\bibitem [{\citenamefont {Hrabec}\ \emph {et~al.}(2014)\citenamefont {Hrabec},
		\citenamefont {Porter}, \citenamefont {Wells}, \citenamefont {Benitez},
		\citenamefont {Burnell}, \citenamefont {McVitie}, \citenamefont {McGrouther},
		\citenamefont {Moore},\ and\ \citenamefont {Marrows}}]{Hrabec2014}%
	\BibitemOpen
	\bibfield  {author} {\bibinfo {author} {\bibfnamefont {A.}~\bibnamefont
			{Hrabec}}, \bibinfo {author} {\bibfnamefont {N.~A.}\ \bibnamefont {Porter}},
		\bibinfo {author} {\bibfnamefont {A.}~\bibnamefont {Wells}}, \bibinfo
		{author} {\bibfnamefont {M.~J.}\ \bibnamefont {Benitez}}, \bibinfo {author}
		{\bibfnamefont {G.}~\bibnamefont {Burnell}}, \bibinfo {author} {\bibfnamefont
			{S.}~\bibnamefont {McVitie}}, \bibinfo {author} {\bibfnamefont
			{D.}~\bibnamefont {McGrouther}}, \bibinfo {author} {\bibfnamefont {T.~A.}\
			\bibnamefont {Moore}}, \ and\ \bibinfo {author} {\bibfnamefont {C.~H.}\
			\bibnamefont {Marrows}},\ }\href {\doibase 10.1103/PhysRevB.90.020402}
	{\bibfield  {journal} {\bibinfo  {journal} {Phys. Rev. B}\ }\textbf {\bibinfo
			{volume} {90}},\ \bibinfo {pages} {020402} (\bibinfo {year}
		{2014})}\BibitemShut {NoStop}%
	\bibitem [{\citenamefont {Nawaoka}\ \emph {et~al.}(2015)\citenamefont
		{Nawaoka}, \citenamefont {Miwa}, \citenamefont {Shiota}, \citenamefont
		{Mizuochi},\ and\ \citenamefont {Suzuki}}]{Nawaoka2015}%
	\BibitemOpen
	\bibfield  {author} {\bibinfo {author} {\bibfnamefont {K.}~\bibnamefont
			{Nawaoka}}, \bibinfo {author} {\bibfnamefont {S.}~\bibnamefont {Miwa}},
		\bibinfo {author} {\bibfnamefont {Y.}~\bibnamefont {Shiota}}, \bibinfo
		{author} {\bibfnamefont {N.}~\bibnamefont {Mizuochi}}, \ and\ \bibinfo
		{author} {\bibfnamefont {Y.}~\bibnamefont {Suzuki}},\ }\href@noop {}
	{\bibfield  {journal} {\bibinfo  {journal} {Appl. Phys. Express}\ }\textbf
		{\bibinfo {volume} {8}},\ \bibinfo {pages} {063004} (\bibinfo {year}
		{2015})}\BibitemShut {NoStop}%
	\bibitem [{\citenamefont {{Yang}}\ \emph {et~al.}()\citenamefont {{Yang}},
		\citenamefont {{Boulle}}, \citenamefont {{Cros}}, \citenamefont {{Fert}},\
		and\ \citenamefont {{Chshiev}}}]{Yang2016d}%
	\BibitemOpen
	\bibfield  {author} {\bibinfo {author} {\bibfnamefont {H.}~\bibnamefont
			{{Yang}}}, \bibinfo {author} {\bibfnamefont {O.}~\bibnamefont {{Boulle}}},
		\bibinfo {author} {\bibfnamefont {V.}~\bibnamefont {{Cros}}}, \bibinfo
		{author} {\bibfnamefont {A.}~\bibnamefont {{Fert}}}, \ and\ \bibinfo {author}
		{\bibfnamefont {M.}~\bibnamefont {{Chshiev}}},\ }\href@noop {} {\ }\Eprint
	{http://arxiv.org/abs/1603.01847 (2016)} {arXiv:1603.01847}
	\BibitemShut {NoStop}%
	\bibitem [{\citenamefont {Shibata}\ \emph {et~al.}(2015)\citenamefont
		{Shibata}, \citenamefont {Iwasaki}, \citenamefont {Kanazawa}, \citenamefont
		{Aizawa}, \citenamefont {Tanigaki}, \citenamefont {Shirai}, \citenamefont
		{Nakajima}, \citenamefont {Kubota}, \citenamefont {Kawasaki}, \citenamefont
		{Park}, \citenamefont {Shindo}, \citenamefont {Nagaosa},\ and\ \citenamefont
		{Tokura}}]{Shibata2015}%
	\BibitemOpen
	\bibfield  {author} {\bibinfo {author} {\bibfnamefont {K.}~\bibnamefont
			{Shibata}}, \bibinfo {author} {\bibfnamefont {J.}~\bibnamefont {Iwasaki}},
		\bibinfo {author} {\bibfnamefont {N.}~\bibnamefont {Kanazawa}}, \bibinfo
		{author} {\bibfnamefont {S.}~\bibnamefont {Aizawa}}, \bibinfo {author}
		{\bibfnamefont {T.}~\bibnamefont {Tanigaki}}, \bibinfo {author}
		{\bibfnamefont {M.}~\bibnamefont {Shirai}}, \bibinfo {author} {\bibfnamefont
			{T.}~\bibnamefont {Nakajima}}, \bibinfo {author} {\bibfnamefont
			{M.}~\bibnamefont {Kubota}}, \bibinfo {author} {\bibfnamefont
			{M.}~\bibnamefont {Kawasaki}}, \bibinfo {author} {\bibfnamefont {H.~S.}\
			\bibnamefont {Park}}, \bibinfo {author} {\bibfnamefont {D.}~\bibnamefont
			{Shindo}}, \bibinfo {author} {\bibfnamefont {N.}~\bibnamefont {Nagaosa}}, \
		and\ \bibinfo {author} {\bibfnamefont {Y.}~\bibnamefont {Tokura}},\ }\href
	{\doibase 10.1038/nnano.2015.113} {\bibfield  {journal} {\bibinfo  {journal}
			{Nat. Nanotechnol.}\ }\textbf {\bibinfo {volume} {10}},\ \bibinfo {pages}
		{589} (\bibinfo {year} {2015})}\BibitemShut {NoStop}%
	\bibitem [{\citenamefont {Chacon}\ \emph {et~al.}(2015)\citenamefont {Chacon},
		\citenamefont {Bauer}, \citenamefont {Adams}, \citenamefont {Rucker},
		\citenamefont {Brandl}, \citenamefont {Georgii}, \citenamefont {Garst},\ and\
		\citenamefont {Pfleiderer}}]{Chacon2015}%
	\BibitemOpen
	\bibfield  {author} {\bibinfo {author} {\bibfnamefont {A.}~\bibnamefont
			{Chacon}}, \bibinfo {author} {\bibfnamefont {A.}~\bibnamefont {Bauer}},
		\bibinfo {author} {\bibfnamefont {T.}~\bibnamefont {Adams}}, \bibinfo
		{author} {\bibfnamefont {F.}~\bibnamefont {Rucker}}, \bibinfo {author}
		{\bibfnamefont {G.}~\bibnamefont {Brandl}}, \bibinfo {author} {\bibfnamefont
			{R.}~\bibnamefont {Georgii}}, \bibinfo {author} {\bibfnamefont
			{M.}~\bibnamefont {Garst}}, \ and\ \bibinfo {author} {\bibfnamefont
			{C.}~\bibnamefont {Pfleiderer}},\ }\href@noop {} {\bibfield  {journal}
		{\bibinfo  {journal} {Phys. Rev. Lett.}\ }\textbf {\bibinfo {volume} {115}},\
		\bibinfo {pages} {267202} (\bibinfo {year} {2015})}\BibitemShut {NoStop}%
	\bibitem [{\citenamefont {Yin}\ \emph {et~al.}(2017)\citenamefont {Yin},
		\citenamefont {Han}, \citenamefont {Kim}, \citenamefont {Lee}, \citenamefont
		{Lee}, \citenamefont {Kim}, \citenamefont {Lee}, \citenamefont {Swagten},\
		and\ \citenamefont {Koopmans}}]{Yin2016}%
	\BibitemOpen
	\bibfield  {author} {\bibinfo {author} {\bibfnamefont {Y.}~\bibnamefont
			{Yin}}, \bibinfo {author} {\bibfnamefont {D.-s.}\ \bibnamefont {Han}},
		\bibinfo {author} {\bibfnamefont {J.-s.}\ \bibnamefont {Kim}}, \bibinfo
		{author} {\bibfnamefont {K.-j.}\ \bibnamefont {Lee}}, \bibinfo {author}
		{\bibfnamefont {S.-w.}\ \bibnamefont {Lee}}, \bibinfo {author} {\bibfnamefont
			{K.-w.}\ \bibnamefont {Kim}}, \bibinfo {author} {\bibfnamefont {H.-w.}\
			\bibnamefont {Lee}}, \bibinfo {author} {\bibfnamefont {H.~J.~M.}\
			\bibnamefont {Swagten}}, \ and\ \bibinfo {author} {\bibfnamefont
			{B.}~\bibnamefont {Koopmans}},\ }\href {\doibase 10.1063/1.4979031}
	{\bibfield  {journal} {\bibinfo  {journal} {Appl. Phys. Lett.}\ }\textbf
		{\bibinfo {volume} {110}},\ \bibinfo {pages} {122401} (\bibinfo {year}
		{2017})}\BibitemShut {NoStop}%
	\bibitem [{\citenamefont {Stenstr\"{o}m}\ \emph {et~al.}(1972)\citenamefont
		{Stenstr\"{o}m}, \citenamefont {Sundstr\"{o}m},\ and\ \citenamefont
		{Sagredo}}]{Stenstrom1972}%
	\BibitemOpen
	\bibfield  {author} {\bibinfo {author} {\bibfnamefont {B.}~\bibnamefont
			{Stenstr\"{o}m}}, \bibinfo {author} {\bibfnamefont {L.~J.}\ \bibnamefont
			{Sundstr\"{o}m}}, \ and\ \bibinfo {author} {\bibfnamefont {V.}~\bibnamefont
			{Sagredo}},\ }\href@noop {} {\bibfield  {journal} {\bibinfo  {journal}
			{Physica Scripta}\ }\textbf {\bibinfo {volume} {6}},\ \bibinfo {pages} {209}
		(\bibinfo {year} {1972})}\BibitemShut {NoStop}%
	\bibitem [{\citenamefont {Petrova}\ \emph {et~al.}(2006)\citenamefont
		{Petrova}, \citenamefont {Bauer}, \citenamefont {Krasnorussky},\ and\
		\citenamefont {Stishov}}]{Petrova2006}%
	\BibitemOpen
	\bibfield  {author} {\bibinfo {author} {\bibfnamefont {A.~E.}\ \bibnamefont
			{Petrova}}, \bibinfo {author} {\bibfnamefont {E.~D.}\ \bibnamefont {Bauer}},
		\bibinfo {author} {\bibfnamefont {V.}~\bibnamefont {Krasnorussky}}, \ and\
		\bibinfo {author} {\bibfnamefont {S.~M.}\ \bibnamefont {Stishov}},\
	}\href@noop {} {\bibfield  {journal} {\bibinfo  {journal} {Phys. Rev. B}\
		}\textbf {\bibinfo {volume} {74}},\ \bibinfo {pages} {092401} (\bibinfo
		{year} {2006})}\BibitemShut {NoStop}%
	\bibitem [{\citenamefont {Leonov}\ \emph
		{et~al.}(2016{\natexlab{b}})\citenamefont {Leonov}, \citenamefont {Loudon},\
		and\ \citenamefont {Bogdanov}}]{Leonov2016}%
	\BibitemOpen
	\bibfield  {author} {\bibinfo {author} {\bibfnamefont {A.~O.}\ \bibnamefont
			{Leonov}}, \bibinfo {author} {\bibfnamefont {J.~C.}\ \bibnamefont {Loudon}},
		\ and\ \bibinfo {author} {\bibfnamefont {A.}~\bibnamefont {Bogdanov}},\
	}\href {\doibase 10.1063/1.4965981} {\bibfield  {journal} {\bibinfo
			{journal} {Appl. Phys. Lett.}\ }\textbf {\bibinfo {volume} {109}},\ \bibinfo
		{pages} {172404} (\bibinfo {year} {2016}{\natexlab{b}})}\BibitemShut
	{NoStop}%
\end{thebibliography}
\end{document}